\documentclass[fleqn,usenatbib]{mnras}

\usepackage{newtxtext,newtxmath}
\usepackage{subcaption}

\usepackage[T1]{fontenc}
\usepackage{subcaption}
\usepackage{ulem}
\usepackage{float}
\usepackage[T1]{fontenc}
\DeclareRobustCommand{\VAN}[3]{#2}
\let\VANthebibliography\thebibliography
\def\thebibliography{\DeclareRobustCommand{\VAN}[3]{##3}\VANthebibliography}

\usepackage{amsmath}	
\usepackage{amssymb}	
\usepackage{graphicx}
\usepackage{graphics}
\usepackage{multirow}

\title[Candidate neutrino blazars]{The broadband spectral energy distribution of candidate neutrino blazars}

\author[Athira M. Bharathan et al.]{
Athira M Bharathan$^{1,2}$\thanks{E-mail: athira.bharathan@res.christuniversity.in},
C. S. Stalin$^{3}$,
Markus Böttcher$^{2}$,
S. Sahayanathan$^{4,5}$,
Blesson Mathew$^{1}$, 
\newauthor
Subir Bhattacharyya$^{4,5}$
\\
$^{1}$Department of Physics and Electronics, CHRIST (Deemed to be University), Bangalore, India\\
$^{2}$North-West University, Potchefstroom, 2531, South Africa\\
$^{3}$Indian Institute of Astrophysics, Block II, Koramangala, Bangalore 560 034, India\\
$^{4}$Astrophysical Sciences Division, Bhabha Atomic Research Centre, Mumbai - 400085, India \\
$^{5}$Homi Bhabha National Institute, Mumbai 400094, India
}

\date{Accepted XXX. Received YYY; in original form ZZZ}

\pubyear{2015}

\begin{document}
\label{firstpage}
\pagerange{\pageref{firstpage}--\pageref{lastpage}}
\maketitle

\begin{abstract}
Blazars, the jet dominated class of AGN comprising flat spectrum radio quasars (FSRQs) and BL Lac objects (BL Lacs) are now increasingly identified as potential sources of high energy neutrinos. Such neutrino blazars are ideal targets to investigate the high energy emission processes and to understand their role as neutrino sources. We report results on four candidate neutrino blazars, PKS 0446+112, TXS 0506+056, PKS 1424$-$418 and PKS 1502+106. We carried out $\gamma$-ray spectral and timing analysis on three time periods that comprise a quiescent epoch, an epoch that corresponds to neutrino detection and a flaring epoch. We also carried out modeling of the  broadband pectral energy distribution (SED) on those three epochs. We found that the $\gamma$-ray spectra of the BL Lac TXS 0506+056 can be adequately described by a power-law, while the spectra of the other three FSRQs require a log-parabola model. On shorter timescales, we observed flux variability with doubling/halving timescales of 4.70 hrs, 9.24 hrs, 30.76 hrs and 15.42 hrs for PKS 0446+112, TXS 0506+056, PKS 1424$-$418 and PKS 1502+106, respectively. The SEDs of most of the epochs for the sources are well explained by a leptonic scenario. However, the quiescent epoch of PKS 1502+106 and the neutrino-emission epoch of PKS 0446+112 required an additional hadronic component to reproduce the observed SEDs. Our analysis reveals a complex interplay of leptonic and hadronic processes. While certain neutrino-associated epochs align with a leptonic model, others necessitate a hadronic component to explain the emission features. 

\end{abstract}



\section{Introduction}

Blazars, which encompass flat-spectrum radio quasars (FSRQs) and BL Lacertae objects (BL Lacs), are a unique subclass of active galactic nuclei (AGN) with their highly relativistic jets aligned closely to the line of sight to the observer \citep{1978PhyS...17..265B}. This separation into FSRQs and BL Lacs is observationally based on the presence of strong  (EW $>$ 5 \AA) or weak (EW $<$ 5 \AA) emission lines in their spectra \citep{1991ApJS...76..813S}. A more physical distinction put forward by \cite{2011MNRAS.414.2674G} is based on the ratio of the luminosity of the broad line region (L$_{BLR}$) to the Eddington luminosity (L$_{Edd}$) with this ratio larger than 5 $\times$ 10$^{-4}$ for FSRQs, while it is lower than  5 $\times$ 10$^{-4}$ for BL Lacs.  The emission from blazars is dominated by their Doppler boosted non-thermal jet emission. They are known to show flux variations across the electromagnetic spectrum from low energy radio to high energy (E > 100 MeV) $\gamma$-rays, and in some cases even very high energy TeV $\gamma$-rays, on a range of time scales from minutes to days to months \citep{1995ARA&A..33..163W, 1997ARA&A..35..445U, 1989Natur.337..627M, 1994BAAS...26..797N, 2004AJ....128...47D, 2018MNRAS.473.1145K, 2021ApJ...923....7B, 2024MNRAS.535.2775V, 2025JHEAp..45...19B}.  In addition to flux variations, they are highly polarized \citep{1980ARA&A..18..321A,2024JApA...45...35B,2024ApJ...975..185B} and are also known to show polarization variations in the radio \citep{2008PASJ...60..707F}, optical \citep{2017ApJ...835..275R,2022MNRAS.517.3236R,2022MNRAS.510.1809P}  and X-ray energies \citep{2024ApJ...963L..41H,2024A&A...689A.119K}.

Blazars exhibit distinctive spectral energy distributions (SED). When represented in the usual log$\nu f(\nu)$ versus log$\nu$ format, the SED displays two prominent broad humps. The low energy component is observed to peak in the infrared-optical energy range (and in some cases in the X-ray energy range), which is attributed to synchrotron emission from relativistic electrons spiraling around the magnetic field lines in the jet. Based on the frequency of the synchrotron peak ($\nu_{syn}$), blazars are further subdivided into three different classes: high, intermediate, and low synchrotron peaked (HSP, ISP and LSP, respectively) blazars, depending on whether $\nu_{syn}$ $>$ 10$^{15}$ Hz, 10$^{14}$ Hz $<$ $\nu_{syn}$ $<$ 10$^{15}$ Hz or $\nu_{syn}$ $<$ 10$^{14}$ Hz \citep{2010ApJ...716...30A}. While the low energy hump is well understood to be of electron synchrotron origin as evidenced through the observed high optical and radio polarization \citep{1997A&A...325..109S,2007ApJ...663..118S}, the origin of the high energy component that peaks in the MeV $-$ GeV energy range is debated between leptonic and hadronic processes. 

In the leptonic scenario, the radiative output across the electromagnetic spectrum is primarily attributed to leptons (electrons and positrons), while any protons in the jet are assumed not to be accelerated to energies high enough to significantly impact the radiative output. In this process, the high-energy emission is produced through inverse Compton scattering, wherein the  low-energy photons are Compton up scattered by the same relativistic electrons that produce synchrotron emission at lower energies. The seed photons could be internal to the jet,  i.e., the synchrotron photons from the jet through a process called synchrotron self Compton (SSC; \citealt{1981ApJ...243..700K,1985ApJ...298..114M,1989ApJ...340..181G}), or photons external to the jet, such as the accretion disk \citep{1993ApJ...416..458D}, the  broad line region \citep{1994ApJ...421..153S,1996MNRAS.280...67G} and the torus \citep{2000ApJ...545..107B,2008MNRAS.387.1669G}, a process called external Compton (EC, \citealt{1987ApJ...322..650B,1989ApJ...340..162M,1992A&A...256L..27D}). In this scenario, one expects to see a close correlation between optical and GeV flux variations since the same population of electrons is responsible for the emission at both the low energy and high energy bands. Such correlations supporting the leptonic scenario are seen in multi-wavelength observations of flaring blazars \citep{2018MNRAS.480.5517L,2019ApJ...880...32L}. However, recent studies concentrating on a systematic investigation of the correlation between optical and GeV flux variations of a large sample of blazars \citep{2019MNRAS.486.1781R,2021MNRAS.504.1772R} have lead to (a) instances of close correlation between optical and GeV flux variations, (b) instances of optical flare without a GeV counterpart and (c) instances of GeV flare without an optical counterpart. These observations (see also \citealt{2023MNRAS.519.6349D}) suggest the high energy component to have contributions from more than one process. 

In contrast to the leptonic process, the hadronic model attributes the high-energy emission to processes involving relativistic protons. In this scenario, the emission can result from synchrotron radiation produced by  protons as they spiral in the magnetic field, or from interactions that lead to photon-pion production \citep{1992A&A...253L..21M,1993A&A...269...67M,2013ApJ...768...54B}. These interactions may result in the generation of very high-energy $\gamma$-rays and secondary particles such as neutrinos, which are unique signatures of hadronic processes. In this scenario, the correlation between optical $-$ GeV flux variations may not be expected.  The recent detection of neutrinos coincident with flaring blazars such as TXS 0506+056 \citep{2018MNRAS.480..192P}, 3HSP J095507.9+355101 \citep{2020A&A...640L...4G} suggests that blazar jets are possible emitters of high energy neutrinos \citep{2020MNRAS.497..865G} and the high energy component in blazar SEDs can have contributions from processes that also involve hadrons. Current observations show that a very small fraction of AGN (say $<$ 1\%) are neutrino emitters \citep{2023ApJ...954...75A}. While blazar SEDs can be equally well represented by leptonic and hadronic models \citep{2013ApJ...768...54B}, direct proof of hadronic interactions from blazar jet is through observation of neutrinos. Yet another potential way to distinguish between leptonic and hadronic origin of high energy emission from blazars is through observations of X-ray and $\gamma$-ray polarization, which are different for leptonic and hadronic models \citep{2013ApJ...774...18Z,2024ApJ...967...93Z}. 

Observations with IceCube \citep{2017JInst..12P3012A} have revealed several spatial associations between high-energy neutrinos and $\gamma$-ray loud blazars, strengthening the case for hadronic interactions in relativistic jets \citep{2016NatPh..12..807K, 2019ApJ...880..103G, 2020MNRAS.497..865G, 2020ApJ...893..162F, 2022ApJ...934L..38B, 2023MNRAS.519.1396S, 2023arXiv230511263B}. Yet, the relative importance of leptonic and hadronic processes in powering the broadband SEDs of blazars remains a subject of debate. Individual case studies illustrate this ambiguity. For example, the neutrino-associated blazar PKS 0735+178 was found to be well described by a leptonic model during the neutrino detection epoch, even though the neutrino association itself implies that hadronic channels cannot be excluded \citep{2024MNRAS.529.3503B}. This highlights a broader challenge; blazars may not all behave in the same way, and the hadronic contribution can be strongly source and epoch-dependent. A recent population-level analysis by \cite{2024A&A...681A.119R} provides a useful perspective of modeling a large sample of 324 \textit{Fermi}-LAT blazars. They showed that in roughly one-third of their sample of sources, the X-ray spectra favor an additional component from hadronic interactions, whereas in the remaining two-thirds, a leptonic model suffices. The corresponding neutrino output from the population accounts for about 20\% of the diffuse IceCube flux, consistent with current stacking limits. Against this backdrop, our work complements the population results by presenting detailed, time-resolved modeling of four blazars associated with IceCube neutrinos. By focusing on multi-epoch SEDs, we explore whether the relative leptonic–hadronic balance changes across flux states and how such variability might connect to neutrino production. In this paper, we present the results of our study on four blazars namely, PKS 0446+112, TXS 0506+056, PKS 1424$-$418 and PKS 1502+106 that are found to be associated with IceCube neutrinos. The choice of these four blazars is based on the availability of multi-wavelength data. 

PKS 0446+112, is a $\gamma$-ray bright FSRQ (4FGL J0449+1121) located at a redshift of $z$ = 2.153 \citep{2024ApJ...968...81P} and powered by a black hole of mass $\sim$7.9 $\times$ 10$^7$ M$_{\odot}$ \citep{2012ApJ...748...49S}. It is highly luminous in $\gamma$-rays (E $>$ 100 MeV) with a value of 1.21 $\times$ 10$^{48}$ erg s$^{-1}$ \citep{2024MNRAS.528.5990S}. It was reported to be spatially coincident with the IceCube event IC-240105 \citep{2024GCN.35485....1I} with an estimated energy of 1.1 TeV. Its time averaged broadband SED has been modeled with a one zone leptonic scenario, with the X-ray and $\gamma$-ray emission being attributed to inverse Compton processes  \citep{2024MNRAS.528.5990S}.

TXS 0506+056 at a redshift of $z$ = 0.3365 \citep{2018ApJ...854L..32P} is a radio bright and a flaring $\gamma$-ray  blazar \citep{2018MNRAS.480..192P} found to be associated with a 290 TeV  neutrino alert, IceCube-170922A by IceCube on 22 September 2017 \citep{2018Sci...361.1378I}. Analysis of archival data prior to this alert has revealed an excess of neutrinos detected towards the direction of TXS 0506+056 during the years 2014$-$2015. However, during this period, the source did not exhibit increased activity in the radio, optical, or $\gamma$-ray energies \citep{2018Sci...361..147I}. These observations led to identify the $\gamma$-ray blazar TXS 0506+056 as the first extragalactic counterpart to a neutrino event. Also, neutrino events associated with the blazar occurred in 2021$-$2022 and 2022$-$2023. The SED of the source during all the four neutrino detection epochs were well reproduced with a stochastic dissipation model \citep{2024ApJ...962..142W} that also explains the episodic neutrino emission from TXS 0506+056.

PKS 1424$-$418, is a FSRQ at a redshift of $z$ =  1.522 \citep{2019MNRAS.482.3458M}. It is listed in the 4FGL catalog with the identifier 4FGL J1427.9-4206 \citep{Abdollahi_2020}, and is powered by a black hole of mass 4.5 $\times$ 10$^{9}$ M$_{\odot}$ \citep{2004ApJ...602..103F}. It is variable in $\gamma$-rays, optical, infrared \citep{2021MNRAS.501.2504A} and sub-mm \citep{2024A&A...692A.203K}. It has an extended one sided morphology in the radio \citep{2024A&A...681A..69B} and is the  most distant TeV-detected blazar associated with an IceCube alert. It was found to be spatially coincident with the IceCube alert IC 121204, which had an estimated energy of 2 PeV, detected on December 4, 2012 \citep{2014PhRvL.113j1101A}. Broadband SED modeling carried out on the source at two flaring epochs found the emission to be well described by a leptonic scenario \citep{2021MNRAS.501.2504A}. 

PKS 1502+106 (4FGL J1504.4+1029), a FSRQ situated at a redshift of $z$ = 1.8378 \citep{2010MNRAS.405.2302H}, was considered a promising neutrino source candidate in the catalog of \cite{2020PhRvL.124e1103A}. It hosts a black hole of mass 4.4 $\times$ 10$^{9}$ M$_{\odot}$ \citep{2011ApJS..194...45S}. It was found to coincide with the IceCube alert IC 190730A \citep{2019ATel12967....1T}, with an estimated energy of 300 TeV.   While the source is not very bright in $\gamma$-rays during the epoch of neutrino detection, it was bright in the radio band \citep{2019ATel12996....1K}. It is variable in the optical \citep{2008ATel.1661....1M} and $\gamma$-ray energies \citep{2010ApJ...710..810A}. However, no significant excess in $\gamma$-ray emission was observed during the IceCube event (IceCube-190730A). According to \cite{2021ApJ...912...54R}, the emission from the source during its quiescent and flaring states can be described by a hadronic model. A summary of the  sources investigated in this study is given in Table \ref{table-1}.

\begin{table*}
\centering
\setlength{\tabcolsep}{4pt}
\renewcommand{\arraystretch}{1.2}
\caption{The details of blazars associated with neutrino detections selected for detailed analysis. Here, T$_0$ is the time in MJD of the neutrino detection, and Reference points to the article that reported the neutrino detection.}
\label{table-1}
\resizebox{\textwidth}{!}{%
\begin{tabular}{llcrcrlll} 
\hline
Source & Class & RA (h:m:s) & Dec (d:m:s) & $z$ & IceCube Alert & T$_0$ (MJD) &  Energy (TeV) & Reference \\ \hline
PKS 0446+112   & FSRQ   & 04:49:07.6711 & +11:21:28.596   & 2.15 & 240105A  & 60314 &  1.1  & \cite{2024GCN.35485....1I} \\ 
TXS 0506+056   & BL Lac & 05:09:25.9644 & +05:41:35.333   & 0.34 & 170922A  & 58018 & 290   & \cite{2018Sci...361..147I} \\ 
PKS 1424$-$418 & FSRQ   & 14:27:56.2975 & $-$42:06:19.437 & 1.52 & 121204   & 56265 & 2000  & \cite{2014PhRvL.113j1101A} \\
PKS 1502+106   & FSRQ   & 15:04:24.9797 & +10:29:39.198   & 1.84 & 190730A  & 58694 & 300   & \cite{2019ATel12967....1T} \\   \hline
\end{tabular}%
}
\end{table*}

\begin{table}
\centering
\caption{Details of the epochs considered for timing, spectral and SED analysis. Here, Quiet refers to the faint state, Neutrino refers to the epoch when neutrino was detected and Flare refers to the epoch when the source was in a flaring state.}
\label{table-2}
\begin{tabular}{llllll} 
\hline
Source         & Epoch  &  State        & MJD             & MJD     & Duration \\ 
               &        &               & start           & end     &  (days)   \\ \hline
PKS 0446+112   & $E_{1}$     & Quiet         & 55500           & 55600   & 100       \\ 
               & $E_{2}$     & Neutrino      & 60265           & 60365   & 100       \\
               & $E_{3}$     & Flare         & 55250           & 55350   & 100       \\
TXS 0506+056   & $E_{1}$     & Quiet         & 58680           & 59000   & 320       \\ 
               & $E_{2}$     & Neutrino      & 57910           & 58100   & 190       \\
               & $E_{3}$     & Flare         & 58360           & 58520   & 160       \\
PKS 1424$-$418 & $E_{1}$     & Quiet         & 55350           & 55750   & 400       \\
               & $E_{2}$     & Neutrino      & 56265           & 56430   & 165       \\
               & $E_{3}$     & Flare         & 56740           & 56940   & 200       \\
PKS 1502+106   & $E_{1}$     & Quiet         & 55900           & 56060   & 160       \\   
               & $E_{2}$     & Neutrino      & 58594           & 58795   & 200       \\
               & $E_{3}$     & Flare         & 57120           & 57135   & 195       \\ \hline
\end{tabular}
\end{table}

This paper is organized as follows. Section 2 details the multi-wavelength data and reduction procedures, the analysis and results are given in Section 3 followed by the discussion and summary in the final Section.  Throughout this paper we used the cosmological constants $\rm \Omega_M = 0.3$, $\rm \Omega_\Lambda = 0.7$, and $\rm H_0 = 70 ~km ~s^{-1} Mpc^{-1}$.

\section{MULTI WAVELENGTH DATA AND REDUCTION}
We analyzed the publicly available data in the  $\gamma$-ray, X-ray, optical, and ultra-violet (UV) bands from August 2008 to February 2022, spanning nearly 14 years for all the sources except PKS 0446+112 for which we used the data from August 2008 to May 2024 in order to cover  the period of the most recent IceCube alert on 5 January 2024, ensuring any contemporaneous $\gamma$-ray activity is accounted for. Use of such a data set will offer insights into the long-term behaviour of the sources as well as understand the processes that are responsible for their multi-wavelength emission.

\subsection{$\gamma$-ray}

In the $\gamma$-ray band, we used the data from the Large Area Telescope (LAT; \citealt{2009ApJ...697.1071A}) on the \textit{Fermi} Gamma-ray Space Telescope. The  \textit{LAT}, is a pair-conversion telescope sensitive to $\gamma$-ray energies from 20 MeV to over 300 GeV. Operating in scanning mode, \textit{Fermi} covers the entire sky approximately every three hours. For the sources TXS 0506+056,  PKS 1424$-$418 and PKS 1502+106 we used data  from August 2008 to February 2022 (MJD: 54500$-$59500; $\sim$170 months), while for PKS 0446+112 we used data from  August 2008 to May 2024 (MJD: 54500$-$60460; $\sim$190 months).  We followed the recommended selection criteria and analysis cuts for PASS8 data using the {\it Fermi} Science Tools version v10r0p5\footnote{http://fermi.gsfc.nasa.gov/ssc/data/analysis/documentation/}. Photon-like events were filtered with the criteria 'evclass=128, evtype=3,' selecting $\gamma$-rays with energies between 0.1 and 300 GeV from a circular region of interest (ROI) with a 15$^\circ$ radius centered on each source. We used a zenith angle cut of 90$^\circ$ to minimize contamination from Earth's limb. For background models, we used the latest isotropic model "iso$\_$P8R2$\_$SOURCE$\_$V6$\_$v06" and the Galactic diffuse emission model "gll$\_$iem$\_$v06". We considered a source to be detected if  the test statistic (TS) exceeded 4, corresponding to a 2$\sigma$ detection significance \citep{1996ApJ...461..396M}. We generated weekly time binned light curves, and sources with TS $<$ 4 were classified as undetected for that time bin.

\subsection{X-ray}

We utilized data in the 0.3$-$10 keV band  from the \textit{Swift} X-ray Telescope (\textit{XRT}; \citealt{2005SSRv..120..165B}) obtained from the HEASARC archives\footnote{https://heasarc.gsfc.nasa.gov/docs/archive.html}. For each of the sources, we reduced the data collected during the specified period  using default parameter values as recommended by the instrument team. In this work, we employed data acquired in the photon counting mode for both timing  and spectral analysis. We used the recent CALDB files from HEASOFT version 6.30 while running the \textit{xrtpipeline}. To extract source spectra we used a circular region with a 60" radius, while for background spectra we used a 120" radius circular region. To generate the the exposure map we used the \textit{XIMAGE} tool, and to generate the  ancillary response files we used the  \textit{xrtmkarf} task. For spectral fitting in XSPEC, we applied an absorbed \textit{power-law} model, incorporating Galactic neutral hydrogen column densities of 7.33 $\times$ 10$^{20}$ cm$^{-2}$ for PKS 1424$-$418, 2.03 $\times$ 10$^{20}$ cm$^{-2}$ for PKS 1502+106, 1.22 $\times$ 10$^{21}$ cm$^{-2}$ for PKS 0446+112, and 1.15 $\times$ 10$^{21}$ cm$^{-2}$ for TXS 0506+056 \citep{2016A&A...594A.116H}.

\subsection{UV and Optical}

For the analysis of UV and optical data covering the entire observation period, we utilized measurements from the \textit{Swift} Ultraviolet/Optical Telescope (UVOT; \citealt{2005SSRv..120...95R}), a key instrument onboard the \textit{Swift} spacecraft \citep{2004ApJ...611.1005G}. The UVOT observations were performed using a combination of the V, U, UVW1, and UVW2 filters, which have central wavelengths (full width at half maximum) values of 5468 \AA ~(796 \AA), 3465 \AA ~(785 \AA), 2600 \AA ~(693 \AA), and 1928 \AA ~(657 \AA), respectively \citep{2008MNRAS.383..627P}. We processed the UVOT data using the online tools available in the \textit{Swift} data archive, which automatically handles aspects such as data calibration and extraction of photometric measurements. To ensure an accurate representation of the intrinsic emission from the sources, we applied corrections for Galactic extinction to the observed UV and optical fluxes using the corresponding extinction values for each region of interest \citep{2011ApJ...737..103S}. 

\section{Analysis and Results}

\subsection{Multi-wavelength light curve}
\label{sec:ml} 

\begin{figure*}
\begin{center}
\includegraphics[width=1.01\textwidth]{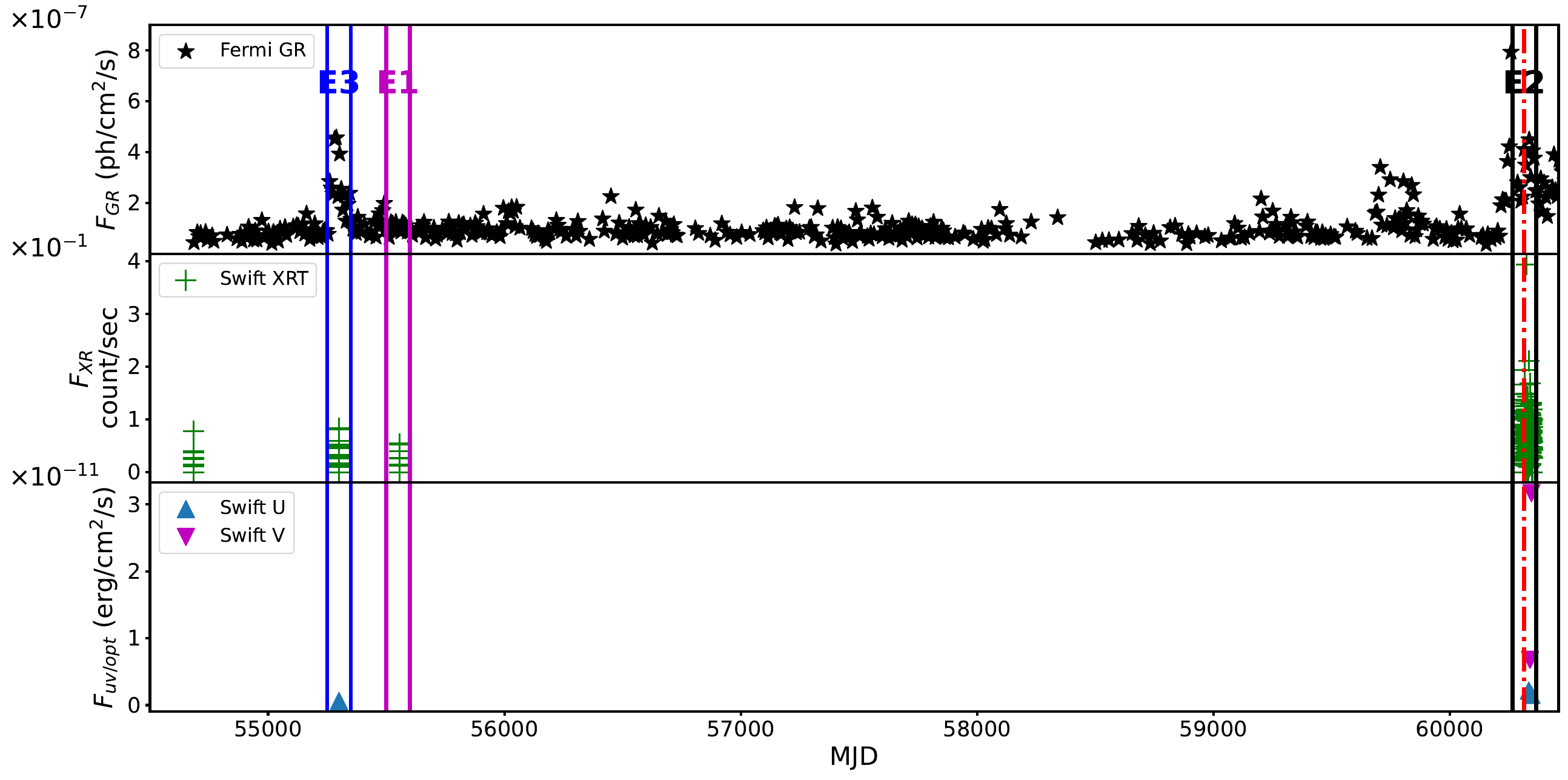}
\includegraphics[width=1.01\textwidth]{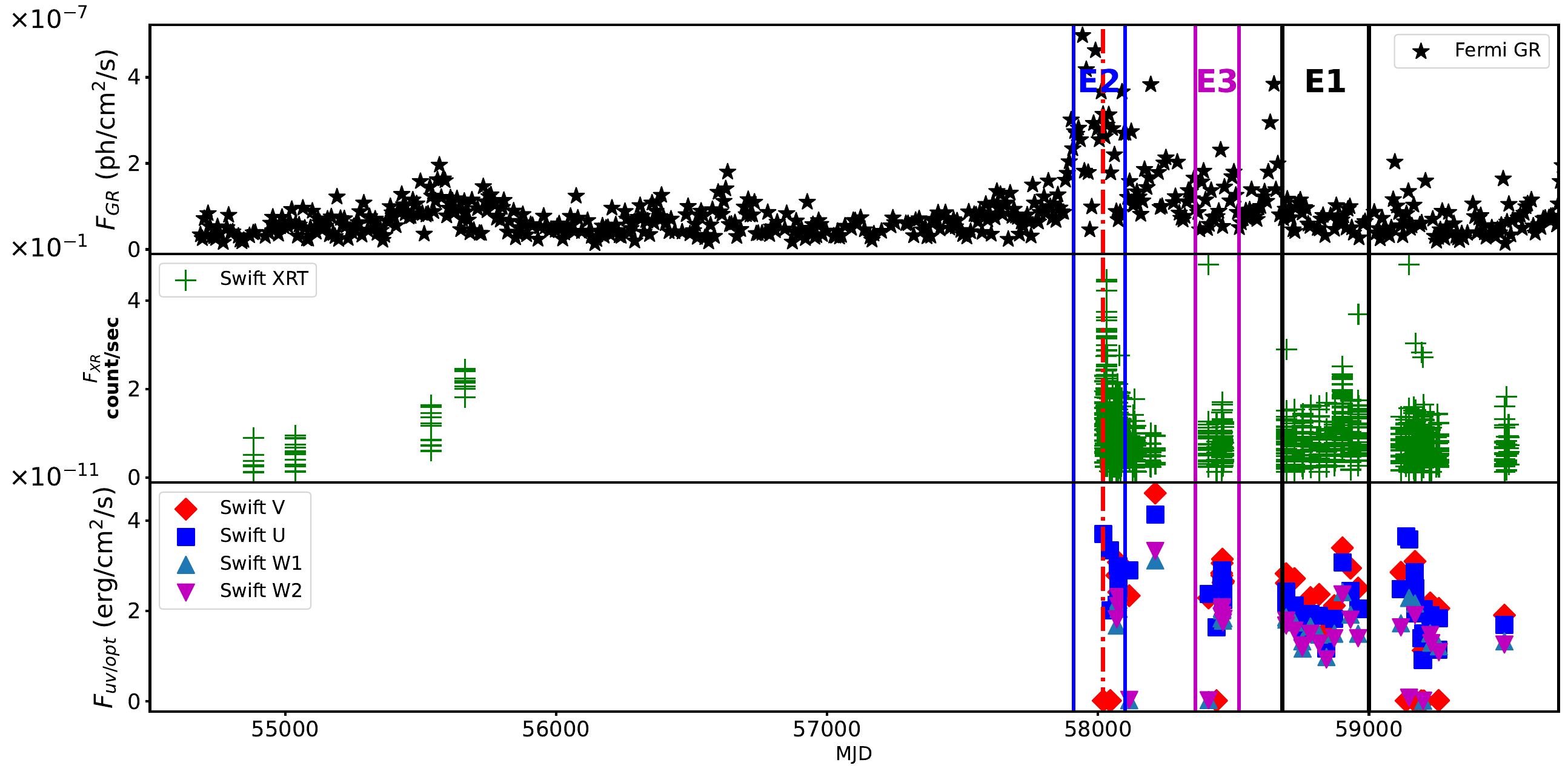}
\end{center}
\vspace*{-0.5cm} \caption{Multi-wavelength light-curves for the sources PKS 0446+112 (top panel) and TXS 0506+056 (bottom panel). In each of the panels, the upper, middle and lower plots show the $\gamma-$ray, X-ray and optical/UV light curves in W1, W2, U and V-filters. The dot-dashed lines refer to the epoch of neutrino detection.}
\label{fig-1}
\end{figure*}

\begin{figure*}
\begin{center}
\includegraphics[width=1.01\textwidth]{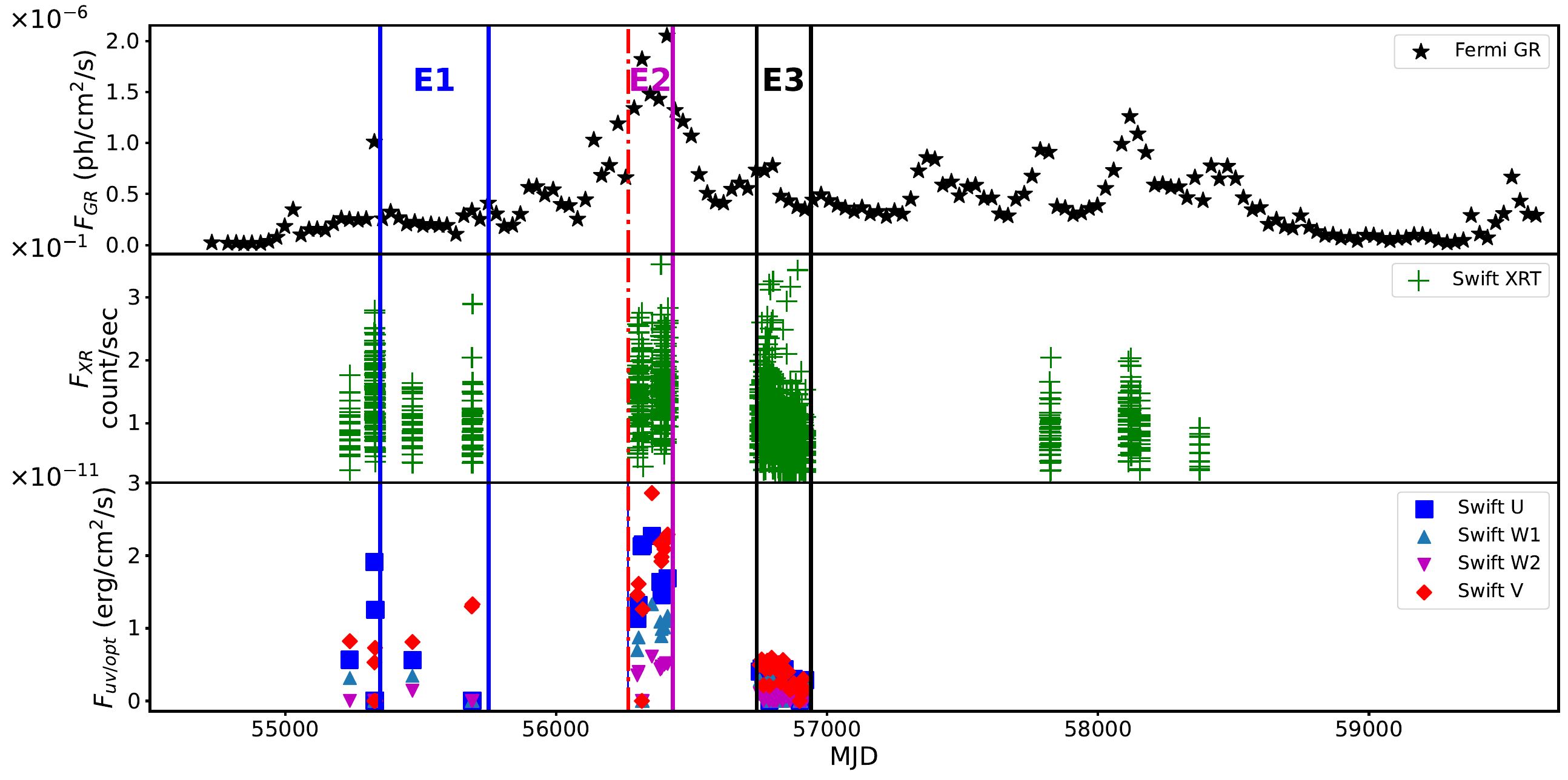}
\includegraphics[width=1.01\textwidth]{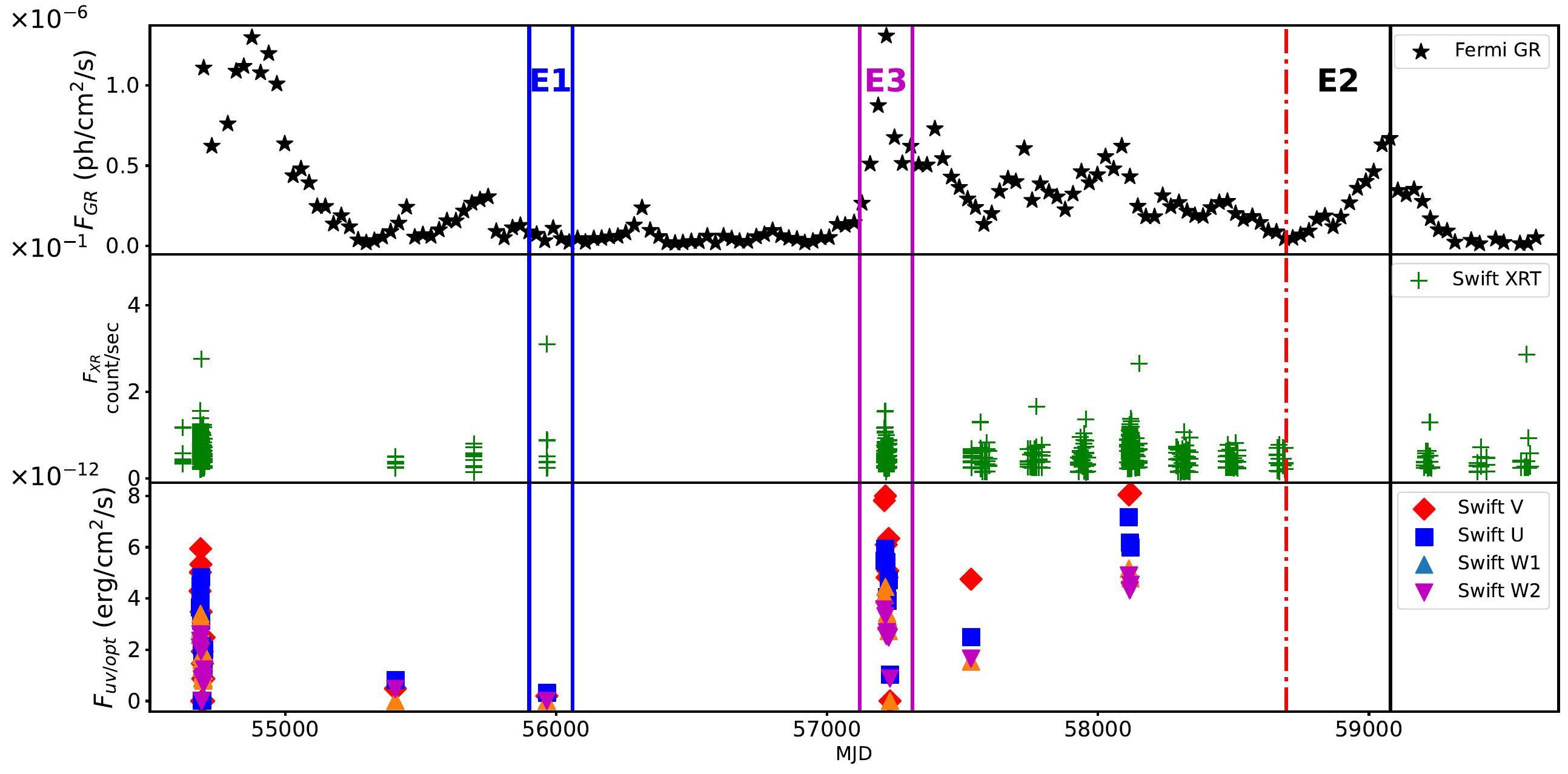}
\end{center}
\vspace*{-0.5cm} \caption{Multi-wavelength light-curves for the sources PKS 1424$-$418 (top panel) and PKS 1502+106 (bottom panel). The other descriptions are same as in Fig. \ref{fig-1}} 
\label{fig-2}
\end{figure*}

The multi-wavelength light curves, that span 14 to 16 years and cover $\gamma$-ray, X-ray, UV, and optical data from August 2008 to February 2022/May 2024 (MJD: 54500–59700/60460), are shown in Fig. \ref{fig-1} for the sources PKS 0446+112 and TXS 0506+056 and in Fig. \ref{fig-2} for the sources PKS 1424$-$418 and PKS 1502+106. From Figs. \ref{fig-1} and \ref{fig-2} it is evident that all the sources showed both low and active states during the period analyzed in this work. For each of the sources, we identified three epochs, denoted as $E_{1}$ (quiescent epoch), $E_{2}$ (epoch of neutrino detection) and $E_{3}$ (flaring epoch) for further analysis. Our definition of quiescent and flaring states in our sample of sources was primarily driven by the availability of near simultaneous multi-band data across wavelengths needed for multi-band SED modeling . They were identified from visual inspection of the long term $\gamma$-ray light curves. The details of these epochs, such as the start time, the end time and the duration are given in Table \ref{table-2}.

The selection of these epochs provides a detailed framework for investigating the spectral and temporal behaviour of these sources during the epoch of neutrino detection relative to other epochs. The source  PKS 0446+112, was in the quiescent  state in the  $\gamma$-ray band most of the time of {\it Fermi} observations, except that it was in a flaring state on two epochs, one at $\sim$MJD 55250 and the other at $\sim$MJD 606265. Notably, it was in its brightest $\gamma$-ray activity state, at the epoch of neutrino emission.  Similarly, a correlation between $\gamma$-ray flaring and neutrino activity was observed for TXS 0506+056, and PKS 1424$-$418 similar to that found in PKS 0735+178 \citep{2024MNRAS.529.3503B}. PKS 1502+106 was in a low $\gamma$-ray activity state during the epoch of neutrino detection (see Fig. \ref{fig-2}). Thus, we found no temporal correlation between $\gamma$-ray flaring activity and neutrino detection in PKS 1502+106.  This was also reported by \cite{2021ApJ...911L..18K}, who noticed that while the source was in a low $\gamma$-ray activity state during the high-energy neutrino detection, the radio emission from PKS 1502+106 was in a high state at the epoch of neutrino detection. Varied correlation between the time period of neutrino detection and $\gamma$-ray flaring activity is also seen in the first neutrino blazar TXS 0506+056. While the $\gamma$-ray flaring activity in TXS 0506+056 during 2017 was found to have a temporal coincidence with the arrival of a 290 TeV neutrino event IC-170922A \citep{2018Sci...361.1378I}, the neutrino detection during 2014/15 \citep{2018Sci...361..147I}  was not accompanied with an increased $\gamma$-ray activity \citep{2020ApJ...893..162F}.

\subsection{Variability analysis}
\subsubsection{Long-term variability}
Long-term variability, spanning weeks to years, offers valuable insights into the processes governing the dynamics of the jet.  To investigate the long-term variability of the sources during the epochs identified in Table \ref{table-2},  we estimated  the fractional variability amplitude in the weekly binned light curves. Following \cite{2003MNRAS.345.1271V} we calculated the fractional variability, $F_{var}$, defined as:
\begin{equation}
\label{eq3}
    F_{var}=\sqrt{\frac{S^2-\overline{\sigma_{err}^2}}{\overline{x}^2}}
\end{equation}
In this equation, $S^2$ represents the variance of the flux over the selected period, $\overline{x}$ is the mean flux over the selected period, and $\overline{\sigma_{err}^2}$ is the mean square of the measurement errors associated with the flux points. We calculated the uncertainty in the fractional variability, $F_{var,err}$, following \cite{2003MNRAS.345.1271V} as 
\begin{equation}
\label{eq4}
    F_{var,err}=\sqrt{\frac{1}{2N}\left(\frac{\overline{\sigma_{err}^2}}{F_{var}\overline{x}^2}\right)^2+\frac{1}{N}\frac{\overline{\sigma_{err}^2}}{\overline{x}^2}}
\end{equation}
Here, $N$ denotes the number of flux points in the light curve. By applying this method, we calculated $\gamma$-ray variability for all the selected epochs in each of the sources analyzed in this work. The results of this variability analysis are presented in Table \ref{table-3}.

\begin{table}
\centering
\caption{Results of $\gamma$-ray variability analysis.}
\begin{tabular}{lll}
\hline
Source & Epoch & $\rm F_{var}(\%)$ \\ \hline
PKS 0446+112   & $E_{1}$  & ---            \\
               & $E_{2}$  & 15.5 $\pm$ 11.8 \\ 
               & $E_{3}$  & 37.1 $\pm$ 5.0 \\ \hline
TXS 0506+056   & $E_{1}$  & ---             \\
               & $E_{2}$  & 41.0 $\pm$ 3.1  \\
               & $E_{3}$  & 33.9 $\pm$ 6.6  \\ \hline
PKS 1424$-$418 & $E_{1}$  & 28.6 $\pm$ 2.1 \\
               & $E_{2}$  & 16.3 $\pm$ 1.1 \\
               & $E_{3}$  & 31.7 $\pm$ 1.7 \\ \hline
PKS 1502+106   & $E_{1}$  & 53.4 $\pm$ 9.4 \\
               & $E_{2}$  & 71.0 $\pm$ 1.9 \\
               & $E_{3}$  & 45.1 $\pm$ 1.5 \\ \hline
\end{tabular}
\label{table-3}
\end{table}

\subsubsection{Short-term variability}
Blazars are known to show flux variations on time scales shorter than an hour \citep{2011A&A...530A..77F,2013ApJ...766L..11S,2022A&A...668A.152P}. Detection of such short time scale variation could enable one to constrain the size and the location of the $\gamma$-ray emission region. We therefore, searched for the presence of flux variations on very short time scales (of the order of hours) in the $\gamma$-ray light curves of  PKS 0446+112, TXS 0506+056, PKS 1424$-$418 and  PKS 1502+106 during their brightest $\gamma$-ray flaring epoch. For that, we  calculated the flux doubling/halving time-scale using the following relation \citep{2013MNRAS.431..824B}
\begin{equation}
F(t) = F(t_0) \times 2^{-(t - t_0)/\tau}
\end{equation}
Here, F(t) and  F(t$_0$) are the fluxes at times t and $t_0$, respectively and $\tau$ is the flux doubling/halving time scale. This calculation was done with the condition that the flux difference between epochs t and t$_0$ is greater than 2$\sigma$ \citep{2011A&A...530A..77F}. Using the flux doubling time scale, we estimated the size of the $\gamma$-ray emitting region as
\begin{equation}
r \leq c\tau\delta/(1+z)
\end{equation}
where, $\delta$ is the Doppler factor,$\tau$ is the flux doubling/halving time-scale and c is the speed of light. The results of the analysis are given in Table \ref{table-4}. The flux doubling/halving time-scale is found to be in hours for BL Lacs and in few days for FSRQs, and the size of the $\gamma$-ray emission region is found to be ranging from (3.7$-$74.7) $\times$ 10$^{15}$ cm. These values are thus within the  range of 1.68$\times$10$^{14}$ cm to 6.61$\times$10$^{19}$ cm found by  \cite{2023ApJS..268...23F} based on an analysis of 2708 {\it Fermi} blazars.

\begin{table}
\centering
\caption{Results of short-term variability during the epoch of very high $\gamma$-ray activity.}
\label{table-4}
\begin{tabular}{lccr} 
\hline
Source         & Epoch & $\tau$ (hr) & $r$ (cm) \\ \hline
PKS 0446+112   & $E_{2}$    &  4.70       &   3.70$\times$10$^{15}$ \\ 
TXS 0506+056   & $E_{2}$    &  9.24       &   7.47$\times$10$^{16}$ \\ 
PKS 1424$-$418 & $E_{2}$    &  30.76      &   2.63$\times$10$^{16}$ \\   
PKS 1502+106   & $E_{3}$    &  15.42      &   7.04$\times$10$^{15}$ \\
\hline
\end{tabular}
\end{table}


\subsubsection{Spectral variability}
\begin{figure*}
\begin{center}
\vbox
   {
    \hbox{
         \includegraphics[width=80mm,height=40mm]{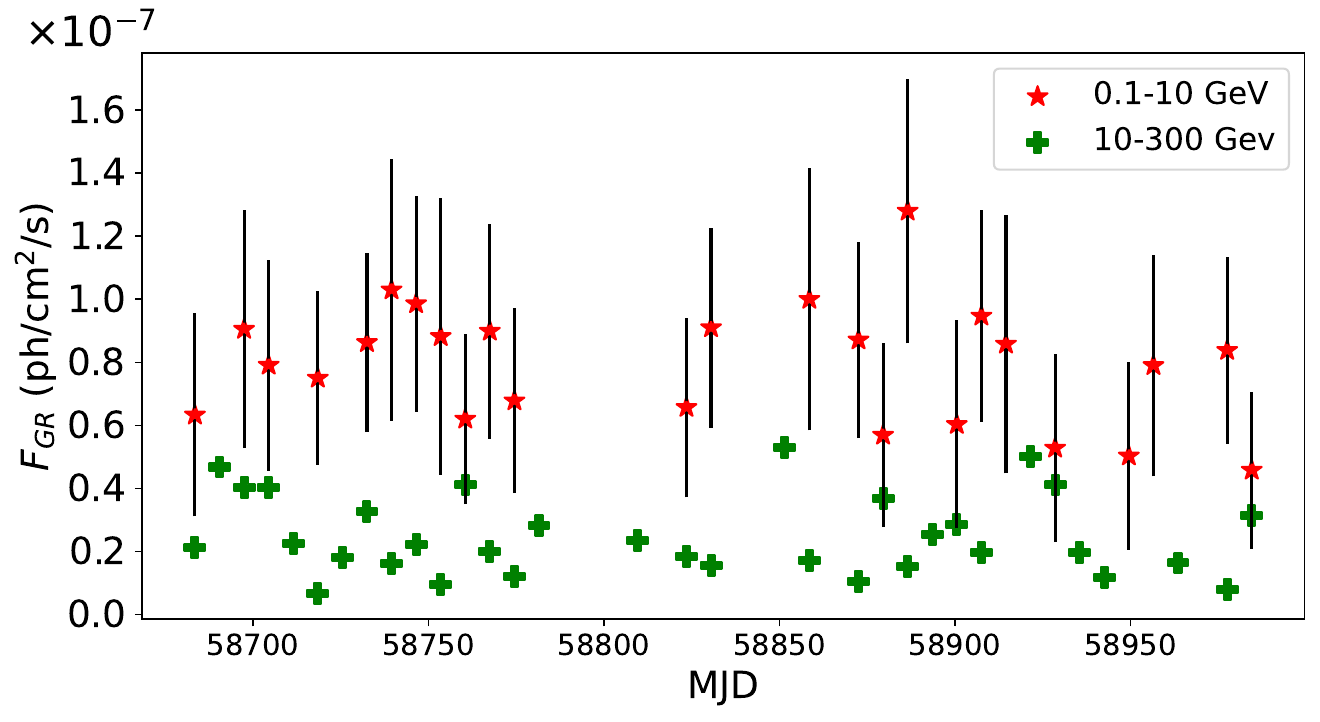}
        \includegraphics[width=80mm,height=40mm]{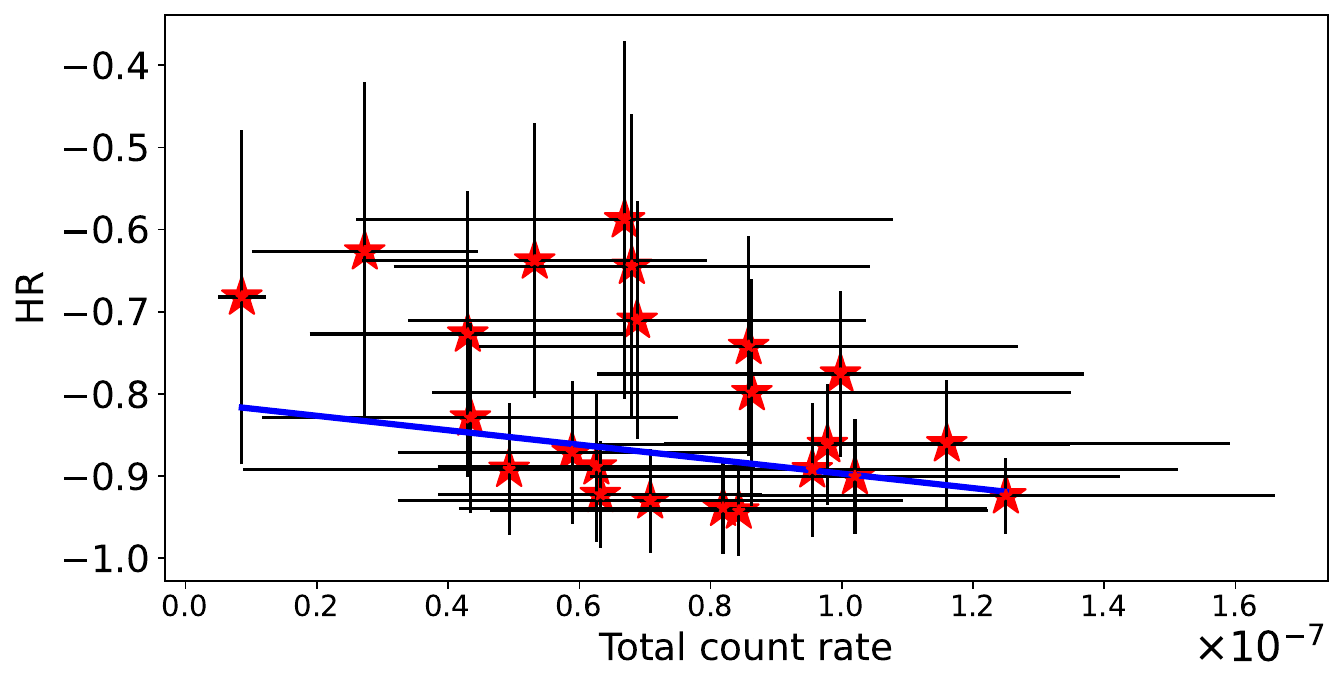}

         }
    \hbox{
         \includegraphics[width=80mm,height=40mm]{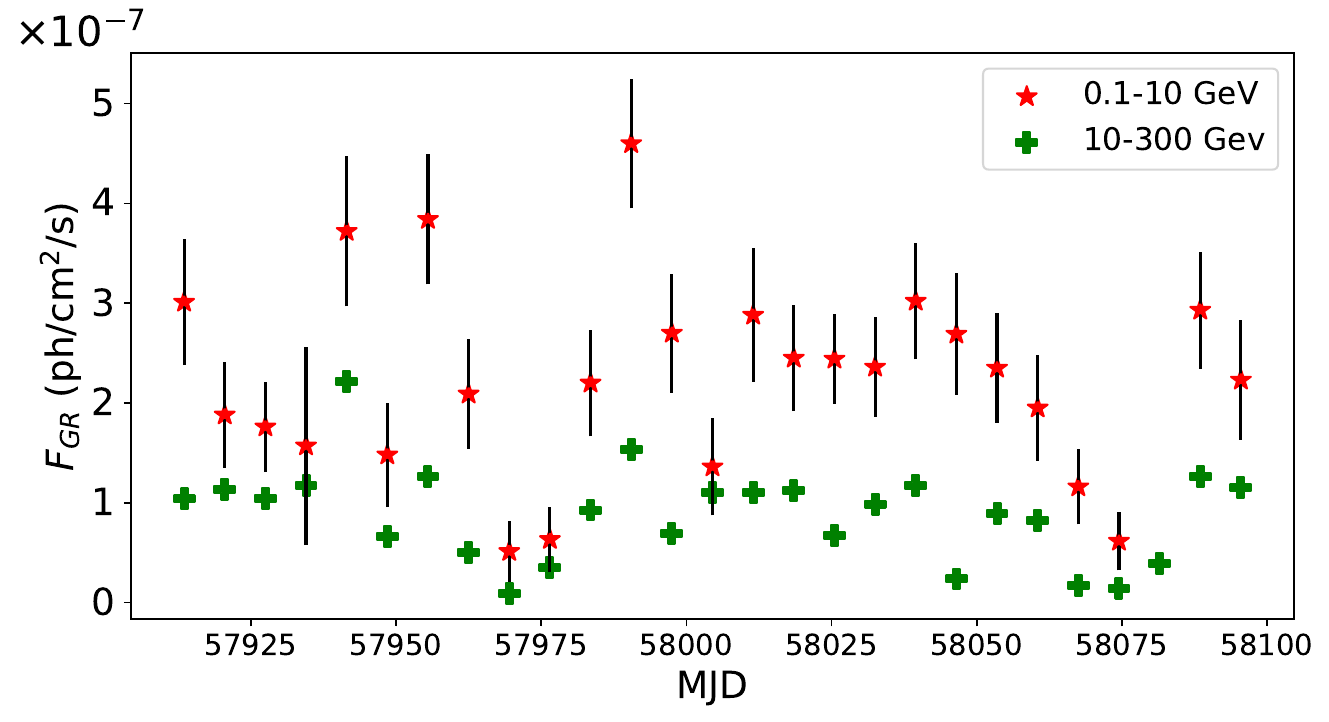}
         \includegraphics[width=80mm,height=40mm]{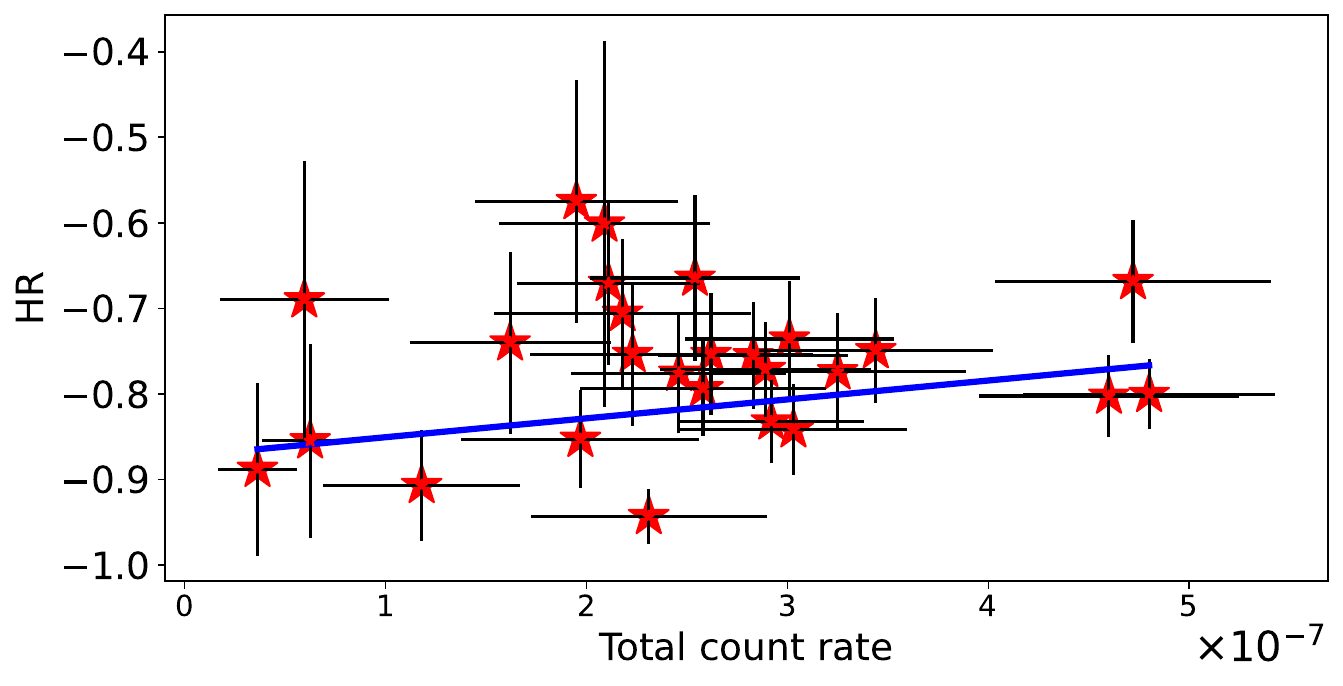}
        }
    \hbox{
         \includegraphics[width=80mm,height=40mm]{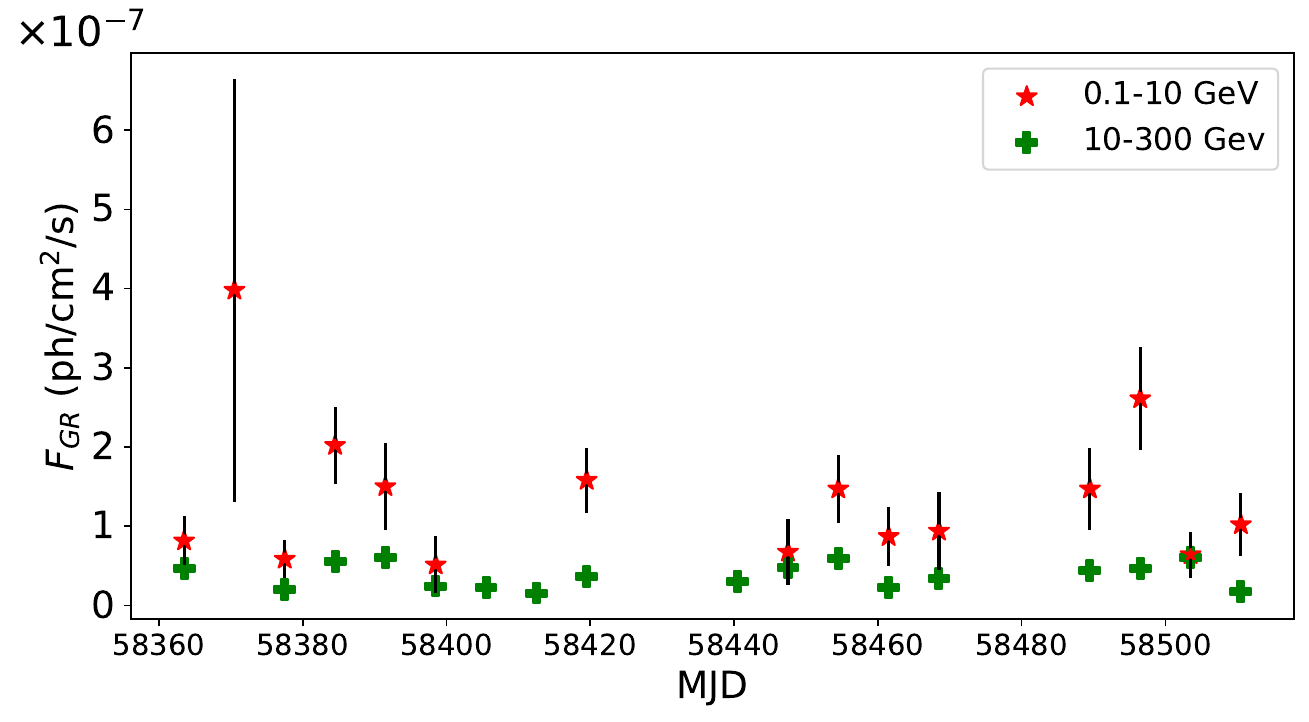}
         \includegraphics[width=80mm,height=40mm]{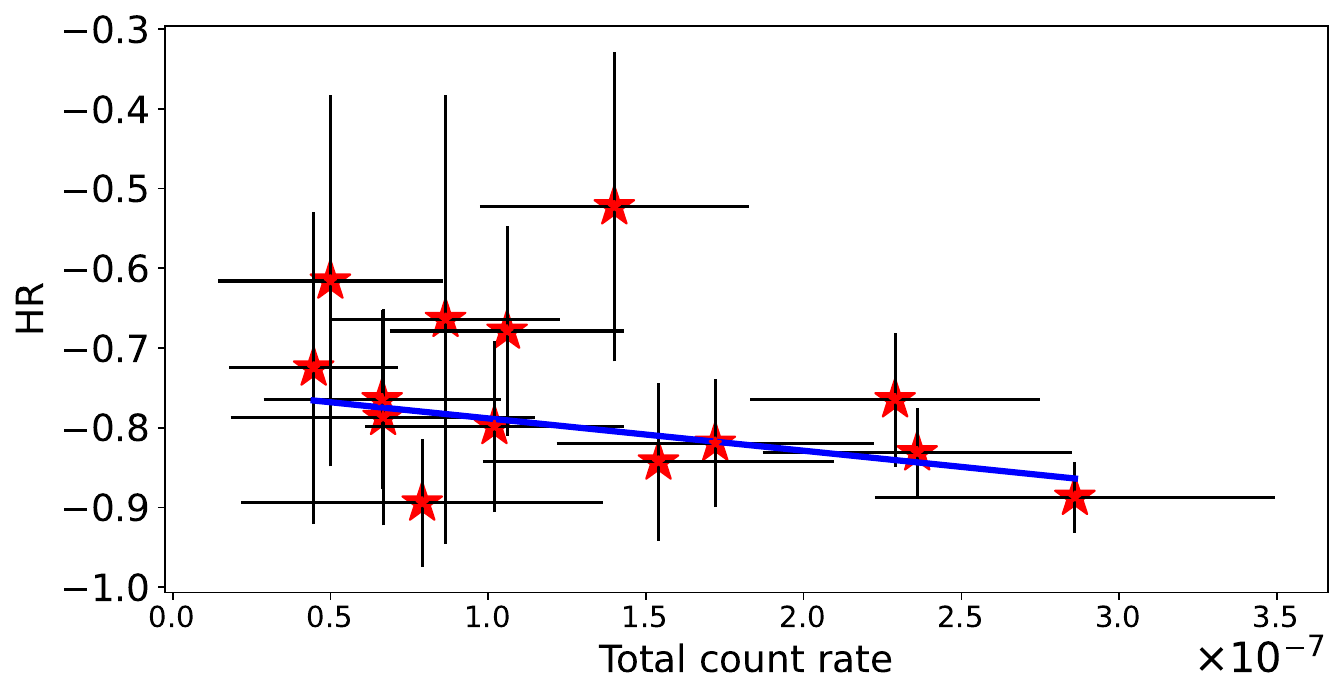}
        }
   } 
\caption{Left panels: One week binned $\gamma$-ray light curves for the epochs $E_{1}$ (top plot), $E_{2}$ (middle plot) and $E_{3}$ (bottom plot) in the 0.1$-$10 GeV (red) and 10$-$300 GeV multiplied by a factor of 5 (green) for TXS 0506+056. Right panels: The HR versus total intensity in the 0.1$-$300 GeV band for the epochs $E_{1}$ (top plot), $E_{2}$ (middle plot) and $E_{3}$ (bottom plot). The solid blue lines are the weighted linear least squares fit to the data.}
\label{fig-3}
\end{center}
\end{figure*}

\begin{figure*}
\begin{center}
\vbox
   {
    \hbox{
         \includegraphics[width=80mm,height=40mm]{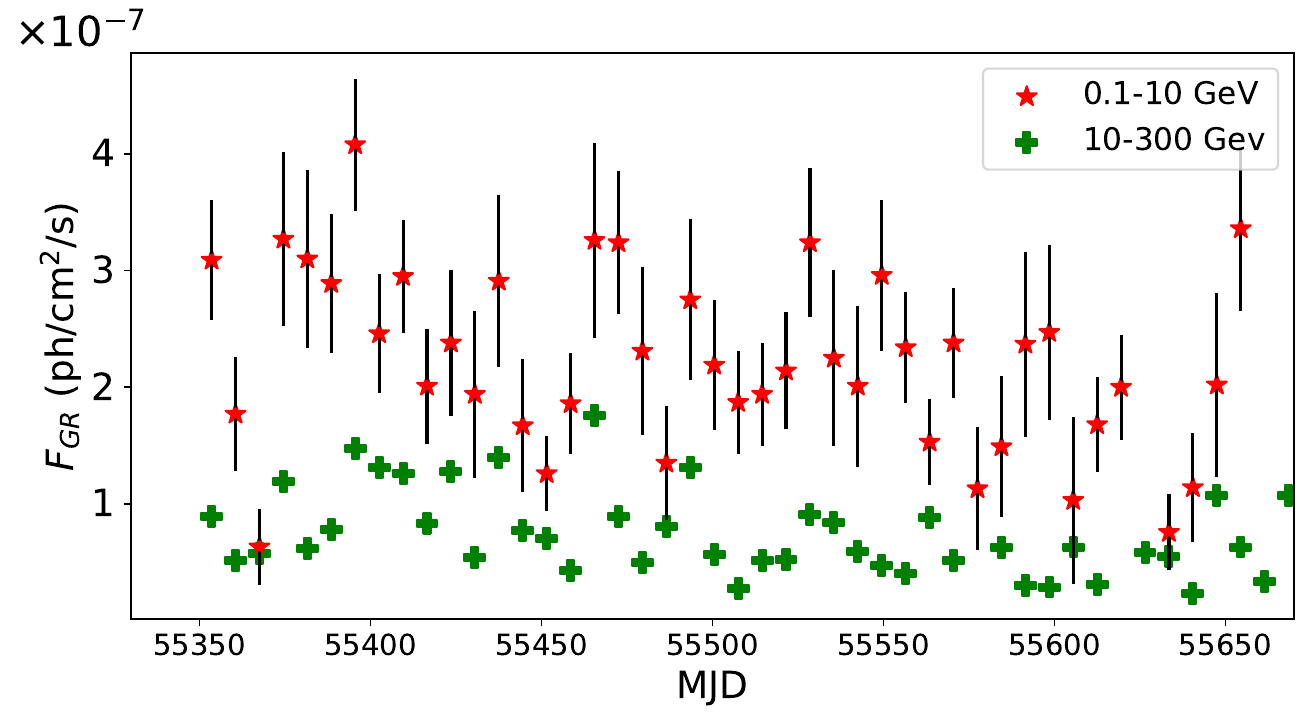}
         \includegraphics[width=80mm,height=40mm]{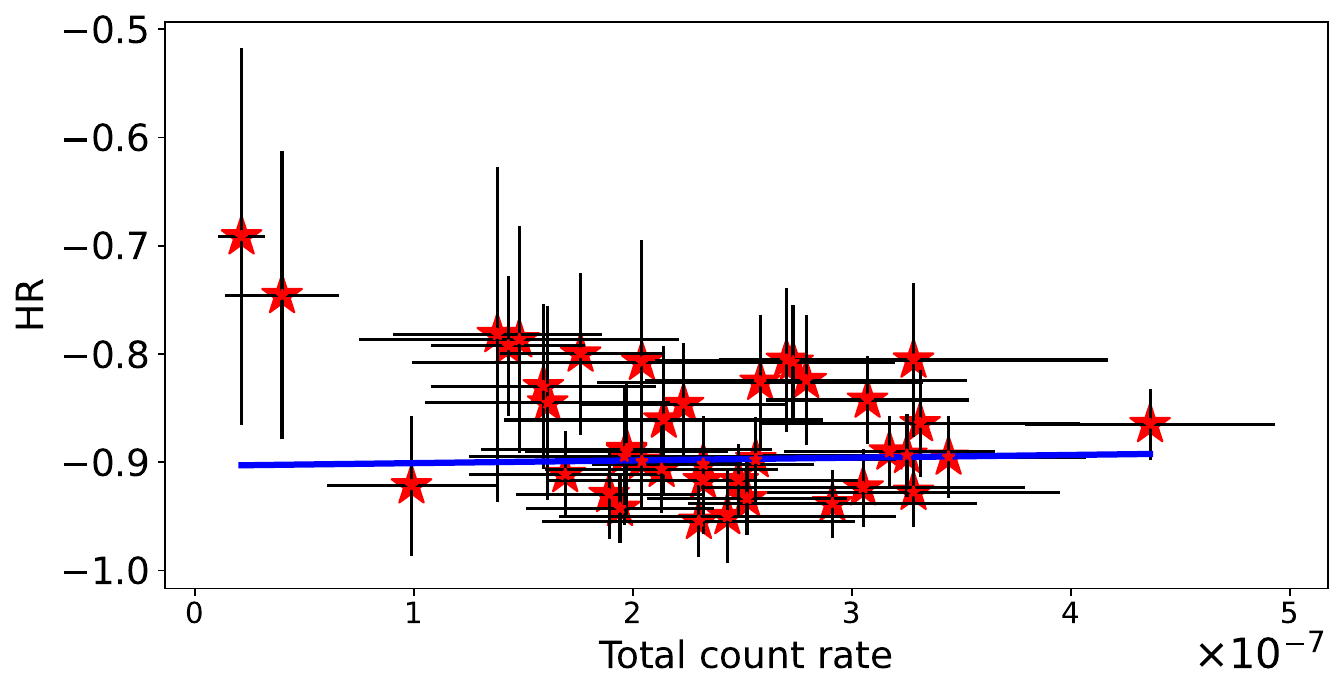}
         }
    \hbox{
         \includegraphics[width=80mm,height=40mm]{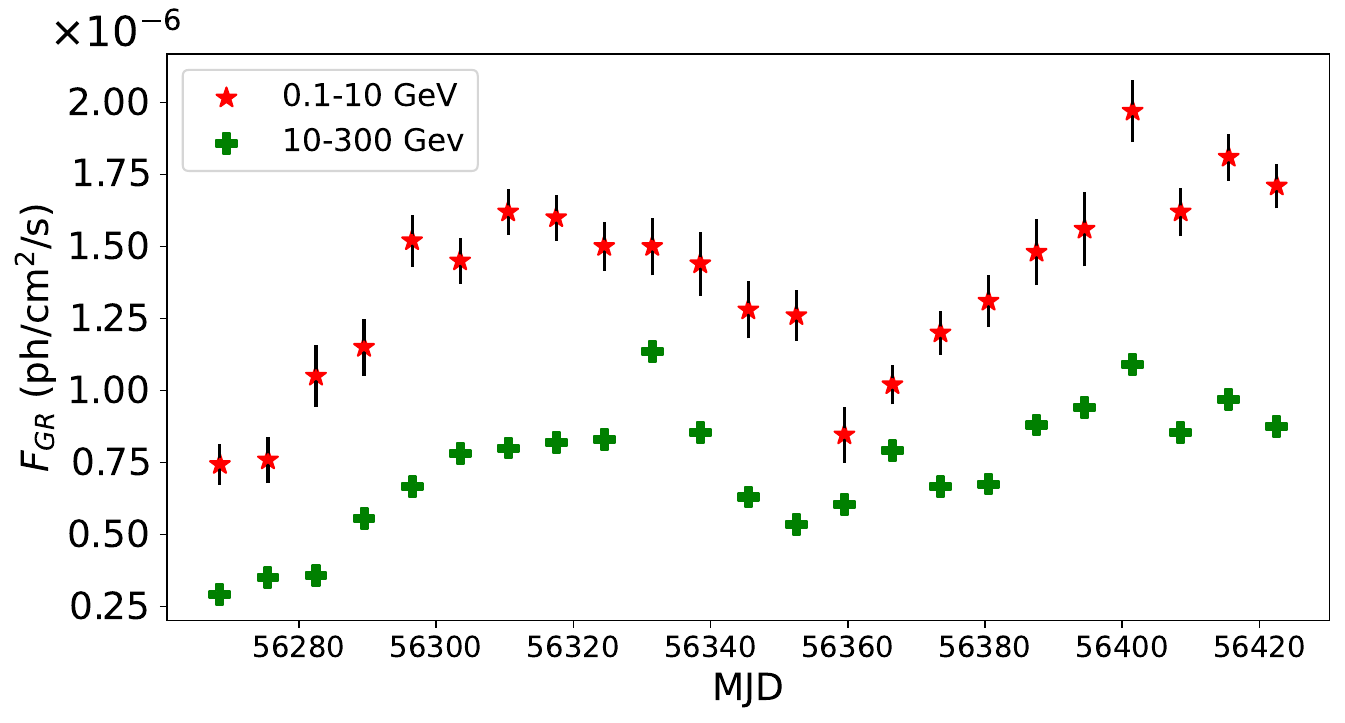}
         \includegraphics[width=80mm,height=40mm]{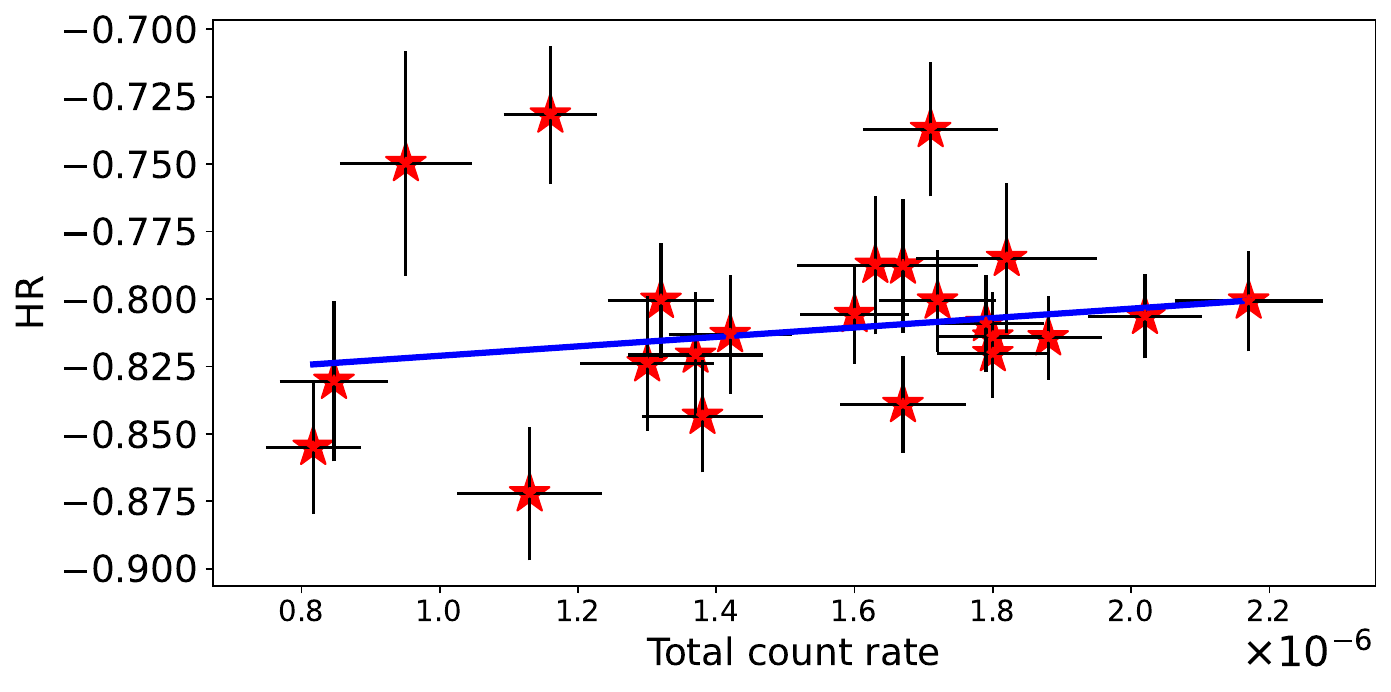}
        }
    \hbox{
         \includegraphics[width=80mm,height=40mm]{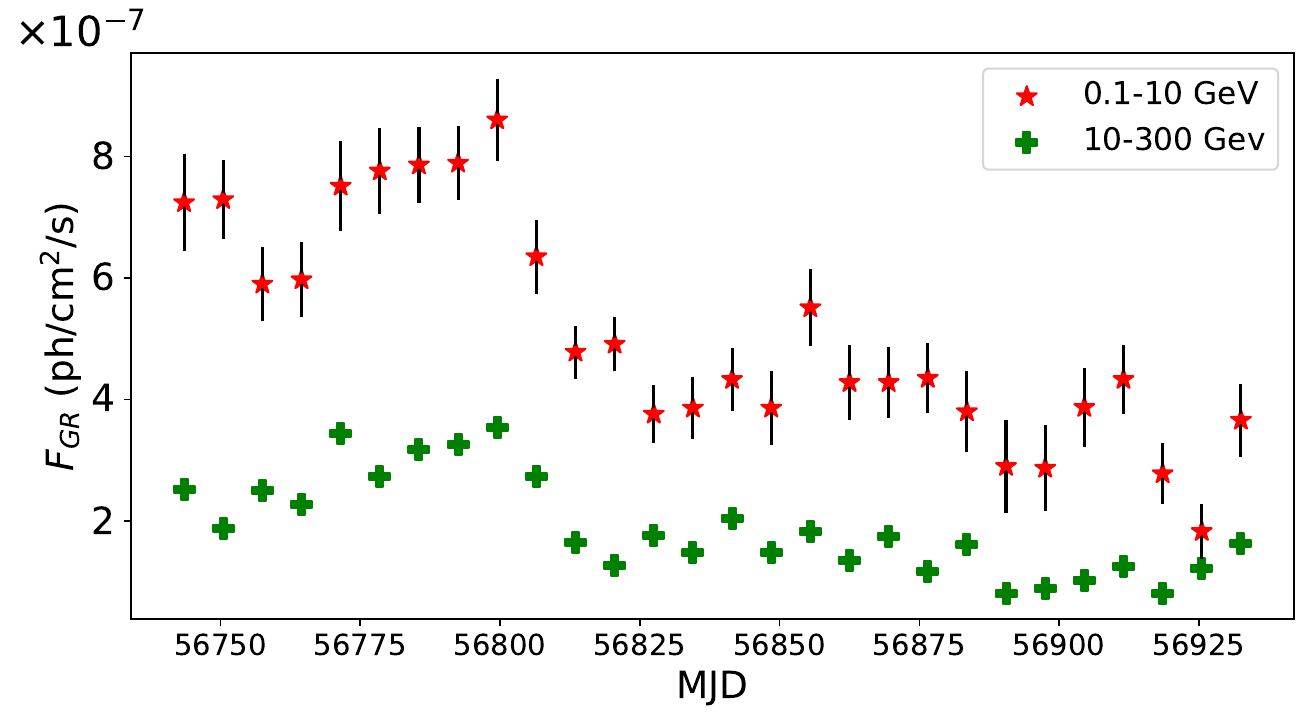}
        \includegraphics[width=80mm,height=40mm]{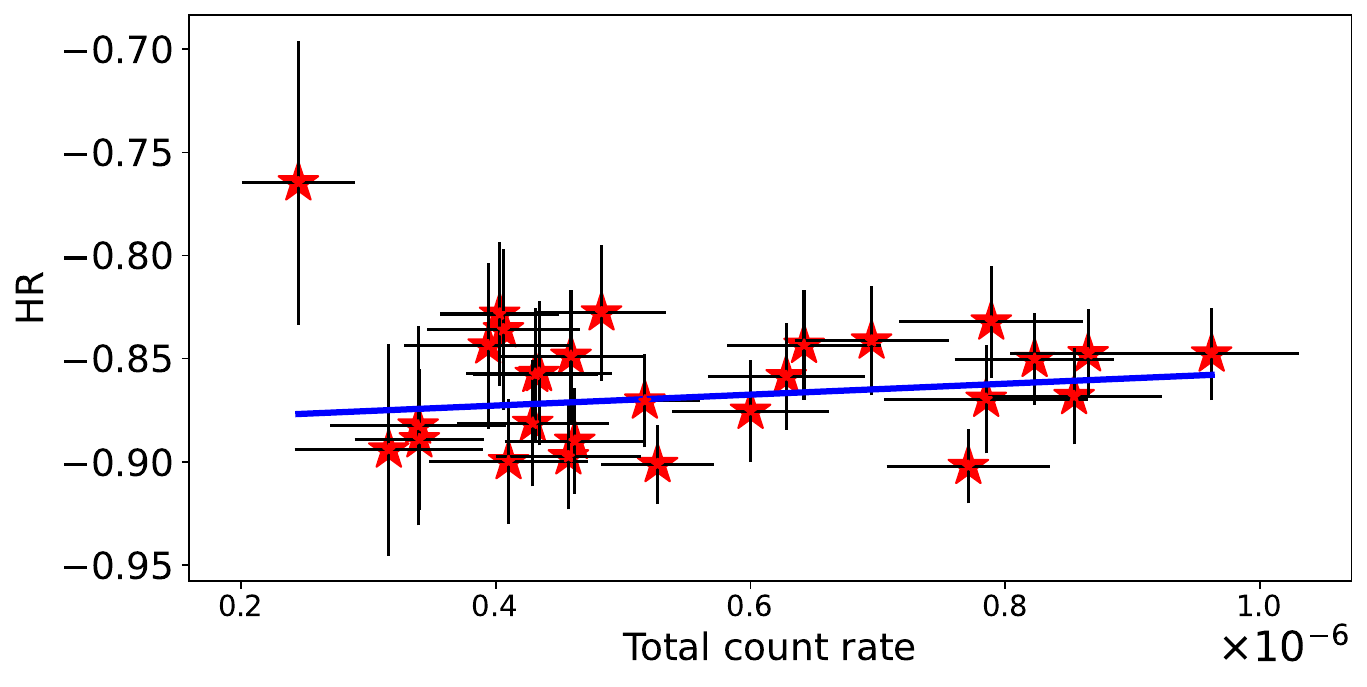}

         }
   } 
\caption{Left panels: One week binned $\gamma$-ray light curves for the epochs $E_{1}$ (top plot), $E_{2}$ (middle plot) and $E_{3}$ (bottom plot) in the 0.1$-$10 GeV (red) and 10$-$300 GeV multiplied by a factor of 3 (green) for PKS 1424$-$418. Right panels: The HR versus total intensity in the 0.1$-$300 GeV band for the epochs $E_{1}$ (top plot), $E_{2}$ (middle plot) and $E_{3}$ (bottom plot).  The solid blue lines are the weighted linear least squares fit to the data.}
\label{fig-4}
\end{center}
\end{figure*}

\begin{figure*}
\begin{center}
\vbox
   {
    \hbox{
         \includegraphics[width=80mm,height=40mm]{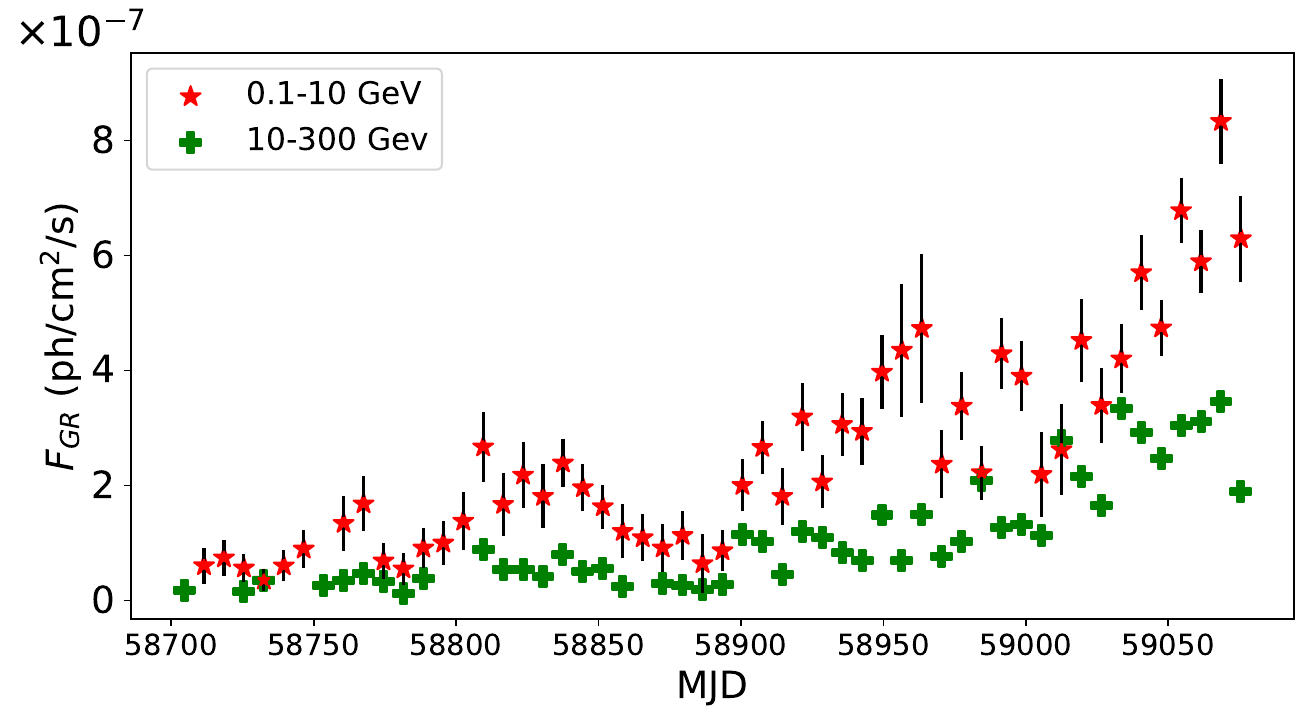}
         \includegraphics[width=80mm,height=40mm]{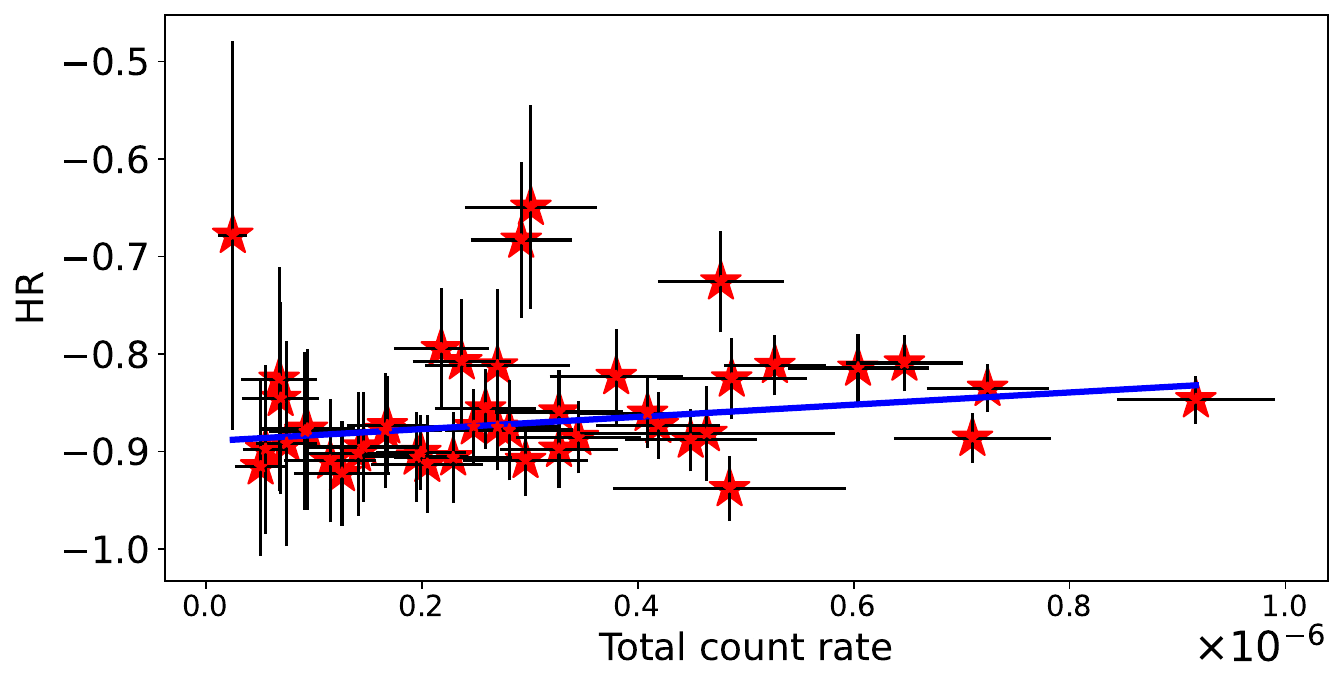}
        }
    \hbox{
         \includegraphics[width=80mm,height=40mm]{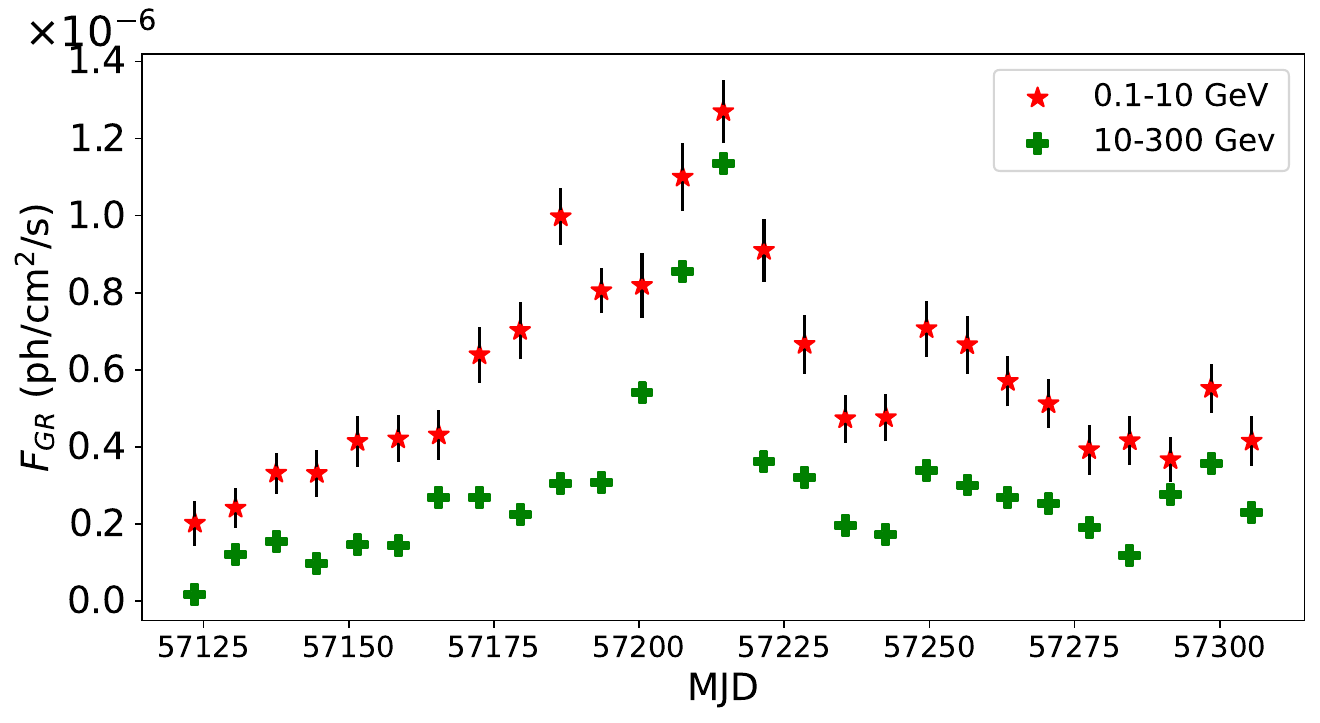}
        \includegraphics[width=80mm,height=40mm]{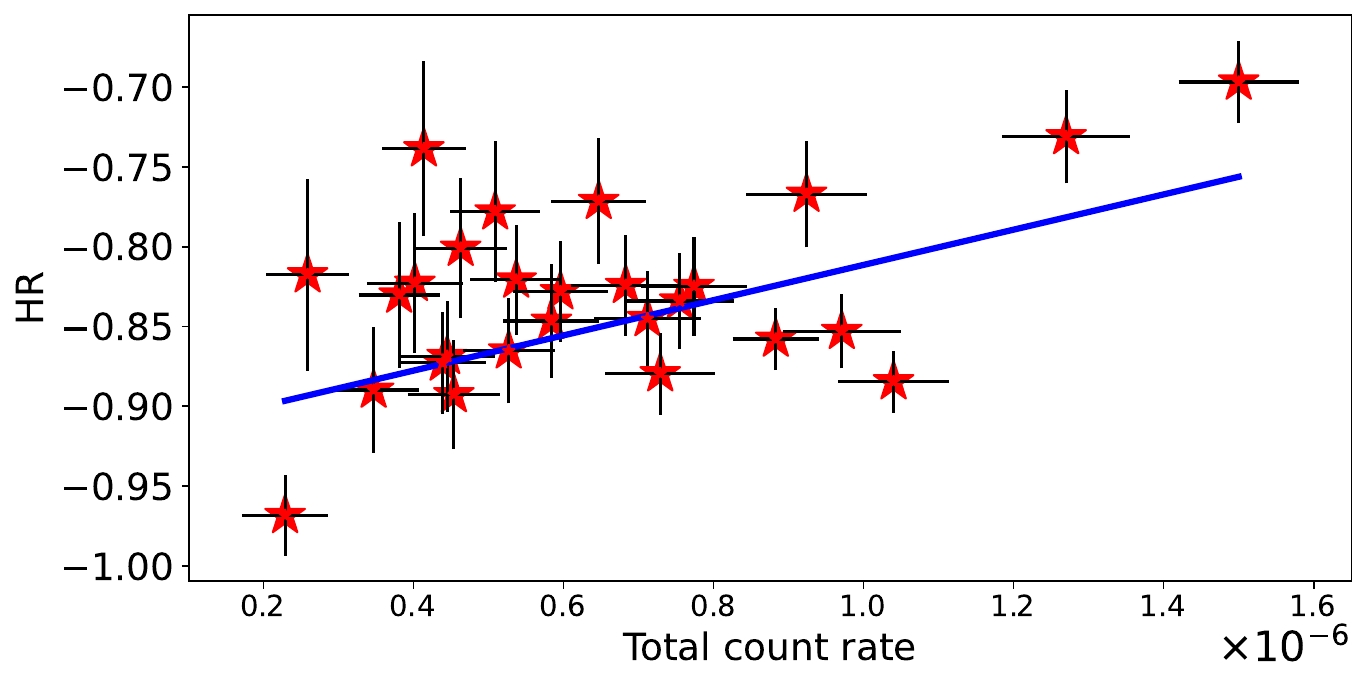}

         }
   } 
\caption{Left panels: One week binned $\gamma$-ray light curves for the epochs $E_{2}$ (top plot), and $E_{3}$ (bottom plot) in the 0.1$-$10 GeV (red) and 10$-$300 GeV multiplied by a factor of 3 (green) for PKS 1502+106. Right panels: The HR versus total intensity in the 0.1$-$300 GeV band for the epochs $E_{2}$ (top plot), and $E_{3}$ (bottom plot).  The solid blue lines are the weighted linear least squares fit to the data.}
\label{fig-5}
\end{center}
\end{figure*}

To examine the spectral variation in the sources during the selected epochs, we employed a model-independent method of calculating the hardness ratio (HR) and analysed its dependence on the source's total flux. For this purpose, we generated  weekly binned $\gamma$-ray light curves in two energy ranges, 0.1$-$10 GeV and 10$-$300 GeV. We then calculated the HR and the associated error as \citep{2008ApJ...678..563J},
\begin{equation}
\label{eq6}
HR= \left(\frac{H-S}{H+S}\right)
\end{equation}

\begin{equation}
\label{eq7}
\sigma_{HR}= \frac{2}{(H+S)^2} \sqrt{H^2\sigma_S^2+S^2\sigma_H^2}
\end{equation}
The light curves in the two energy ranges and the corresponding variation of HR with brightness of the sources for the selected epochs, where HR could be evaluated  are shown in Fig. \ref{fig-3} for TXS 0506+056, Fig. \ref{fig-4} for PKS 1424$-$418 and Fig. \ref{fig-5} for PKS 1502+106.  We also carried out a weighted linear least squares fit to the HR versus intensity in the 0.1 $-$ 300 GeV band and the results of the fit are given in Table \ref{table-5}. For TXS 0506+056, the diagram during epoch $E_{1}$ shows a possible softer-when-brighter behaviour. However, the scatter in the data points is relatively large, and during epochs $E_{2}$ and $E_{3}$ no clear correlation between HR and brightness is evident. For PKS 1424$-$418, the data in epoch $E_{1}$ suggest a harder-when-brighter trend, although the scatter prevents a firm conclusion. No clear trend is observed during epochs $E_{2}$ and $E_{3}$. For PKS 1502+106, the distribution during epoch $E_{3}$ indicates a possible harder-when-brighter behaviour, while during epoch $E_{2}$ no significant correlation is apparent.

\begin{table}
\centering
\caption{Results of the weighted linear least squares fit to the HR v/s total intensity diagram. Here, R is the linear correlation coefficient and p is the 
probability for no correlation. }
\begin{tabular}{llllll}
\hline
Source           & Epochs  & slope      & intercept  & R      & p   \\ \hline
TXS 0506+056     & $E_{1}$      & $-$879573    & $-$0.809     & $-$0.482 & 0.019 \\ 
                 & $E_{2}$      & 220934     & $-$0.873     & 0.124  & 0.546 \\ 
                 & $E_{3}$      & $-$405588    & $-$0.748     & $-$0.388 & 0.170 \\ \hline
PKS 1424$-$418   & $E_{1}$      & 25209      & $-$0.903     & $-$0.403 & 0.008 \\
                 & $E_{2}$     & 17464      & $-$0.839     & 0.154  & 0.480 \\
                 & $E_{3}$      & 26468      & $-$0.883     & $-$0.008 & 0.969 \\ \hline
PKS 1502+106     & $E_{2}$      & 62605      & $-$0.889     & 0.133  & 0.381 \\
                 & $E_{3}$      & 110454     & $-$0.922     & 0.492  & 0.009 \\ \hline     
\end{tabular}
\label{table-5}
\end{table}

\subsection{$\gamma$-ray spectra}
Analysis of the $\gamma$-ray spectra can provide insights into the mechanisms responsible for high-energy emission.  We analyzed the $\gamma$-ray spectra for all the selected epochs in the sources using two spectral models namely the \textit{power law} (PL) and \textit{log parabola} (LP) models. The PL model has the functional form 
\begin{equation}
\label{eq8}
\frac{dN(E)}{dE} = N_0 E^{-\Gamma}
\end{equation}
Here, $\frac{dN(E)}{dE}$ is the differential photon number (cm$^{-2}$ s$^{-1}$ MeV$^{-1}$),  $\rm N_0$ is the normalization and $\Gamma$ is the photon index. 

The LP model has the functional form \citep{2012ApJS..199...31N},
\begin{equation}
\frac{dN(E)}{dE}=N_{\circ}(E/E_{\circ})^{-\alpha-\beta ln(E/E_{\circ})}
\end{equation}
where, $\alpha$ is photon index at $E_{\circ}$, $\beta$ defines the peak spectral curvature in the spectra, N$_{\circ}$ and E$_{\circ}$ are the normalization and scaling factor of the energy spectrum, respectively.

\begin{figure*}
\begin{center}
\vbox{
\hbox{
     \includegraphics[scale=0.27]{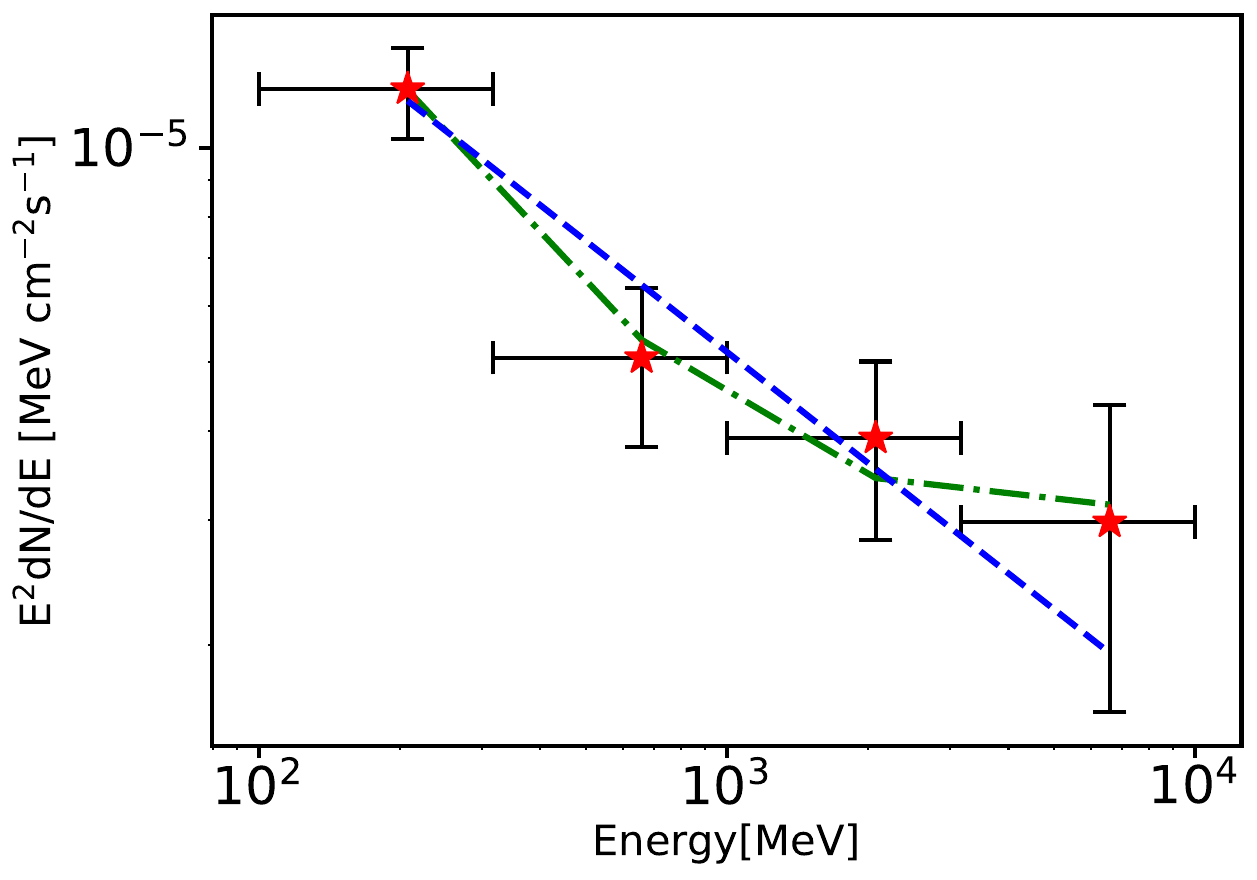}
     \includegraphics[scale=0.27]{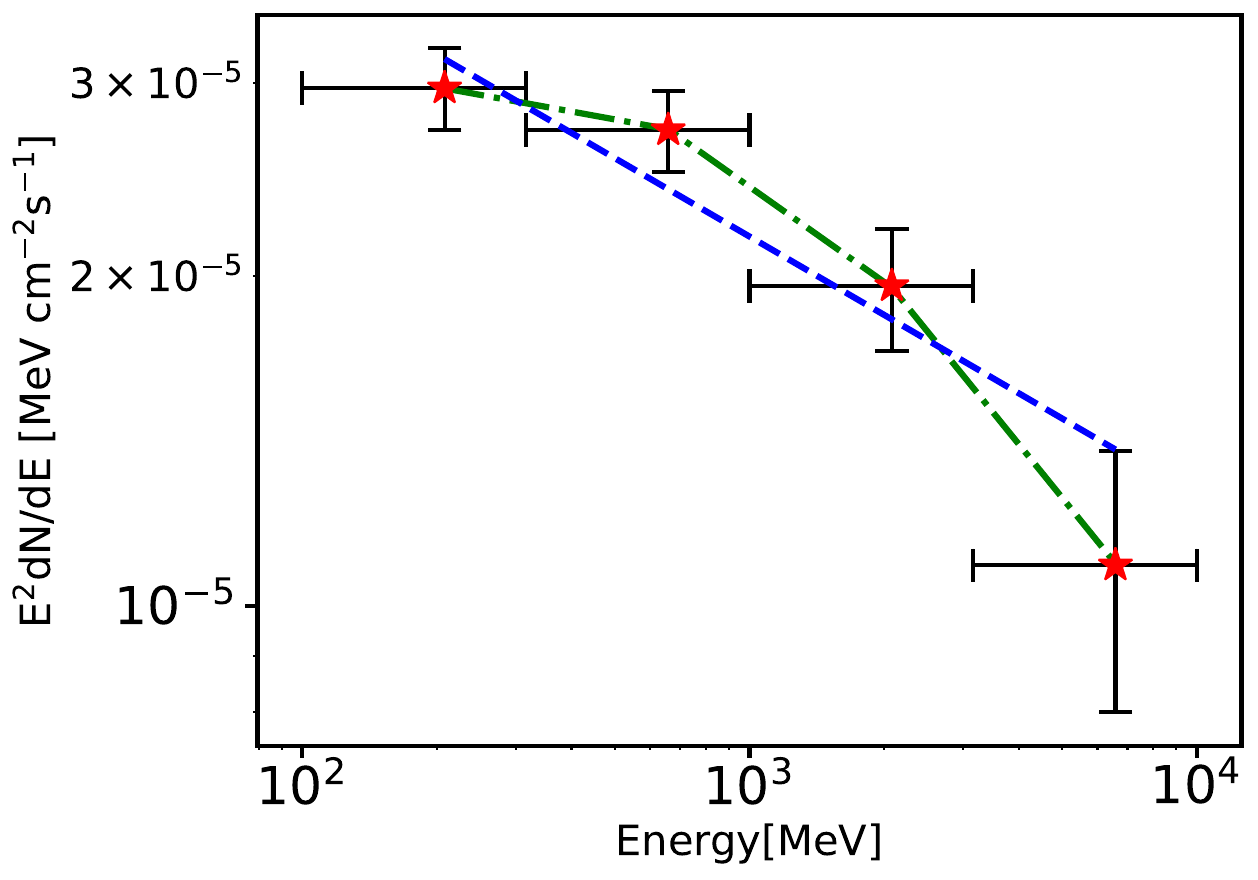}
     \includegraphics[scale=0.27]{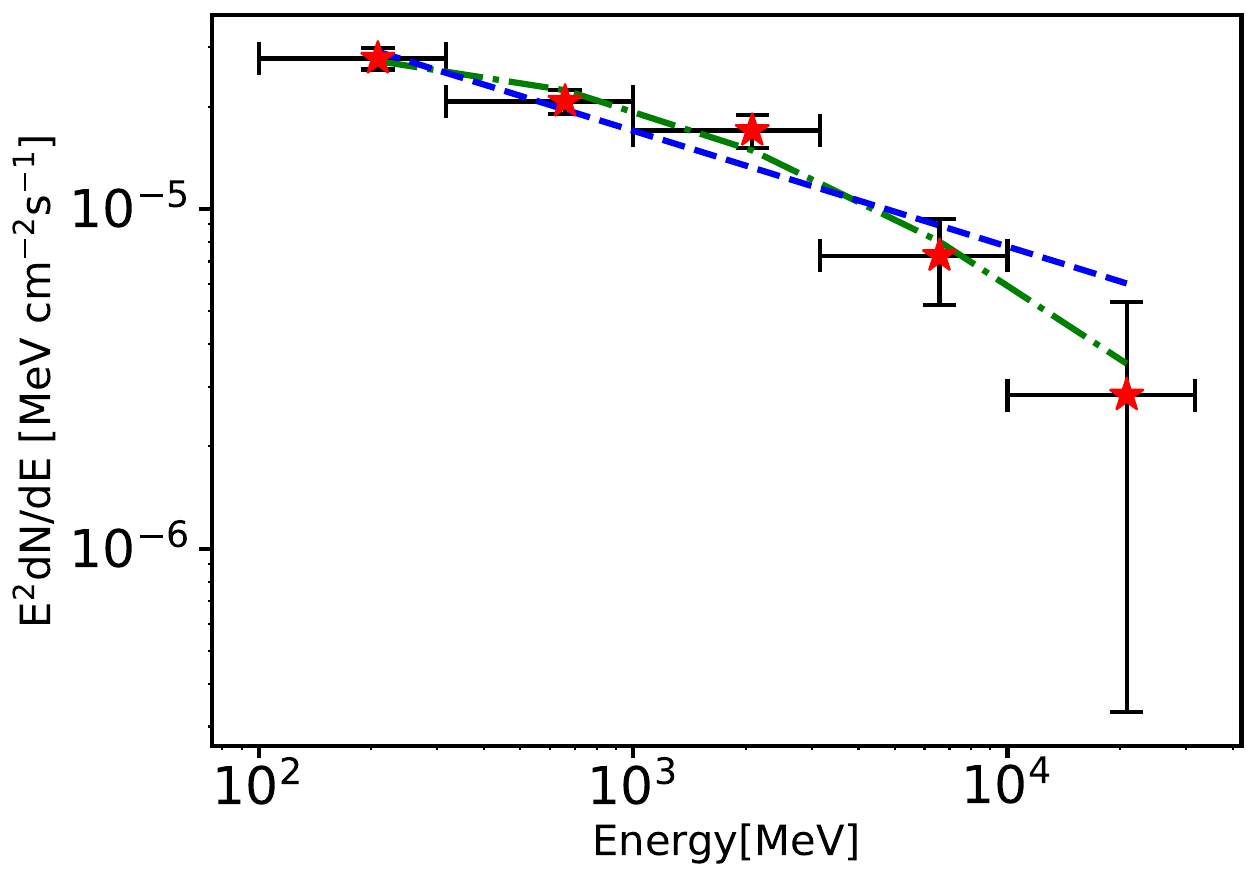}
      }

\hbox{
     \includegraphics[scale=0.27]{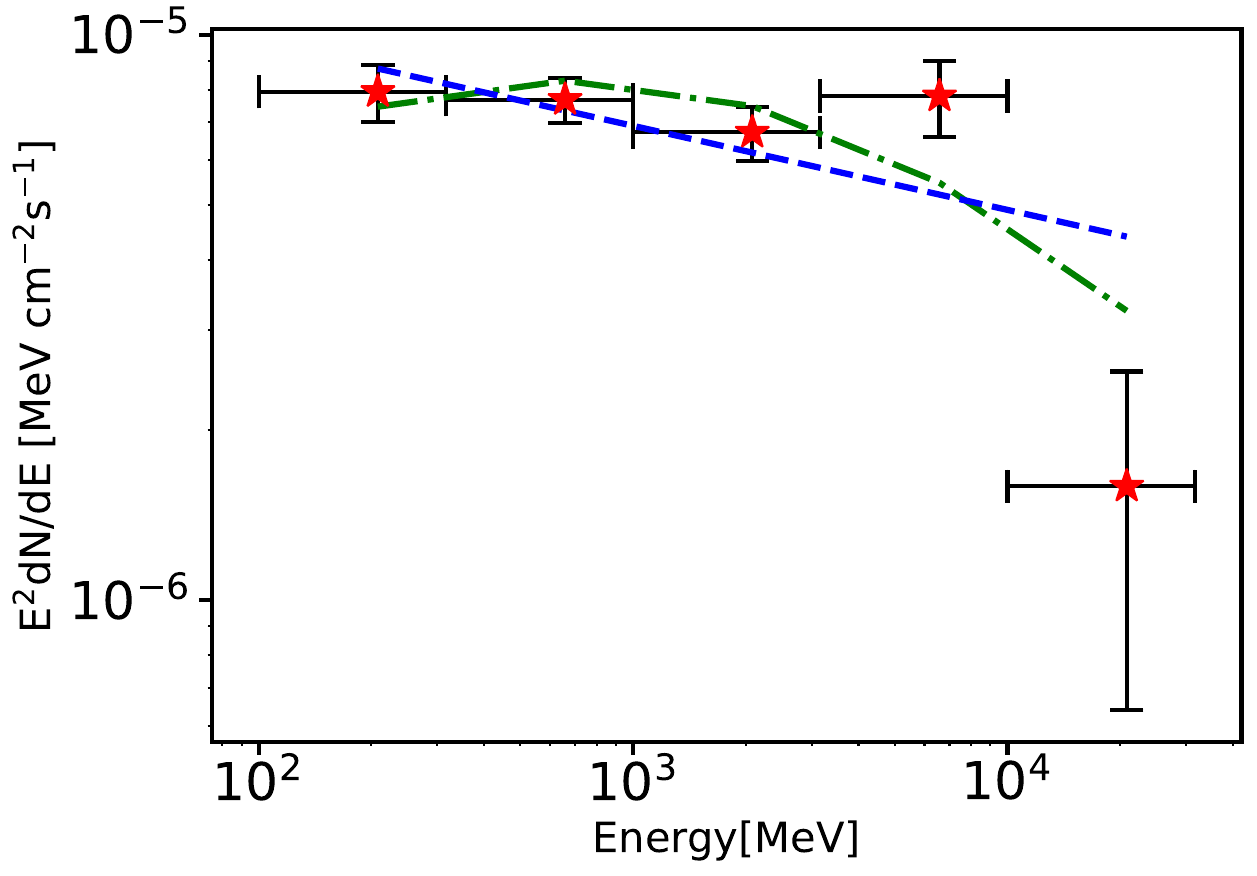}
     \includegraphics[scale=0.27]{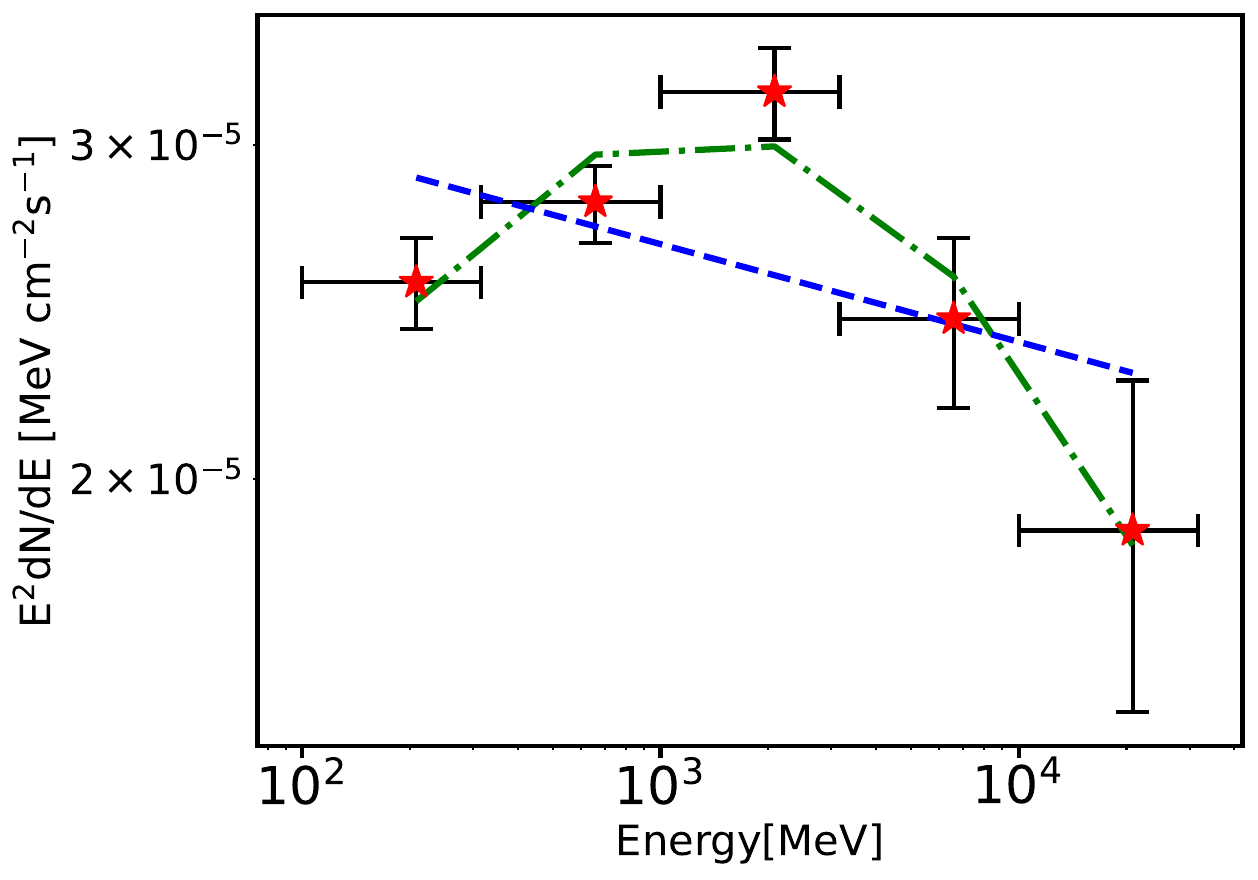}
     \includegraphics[scale=0.27]{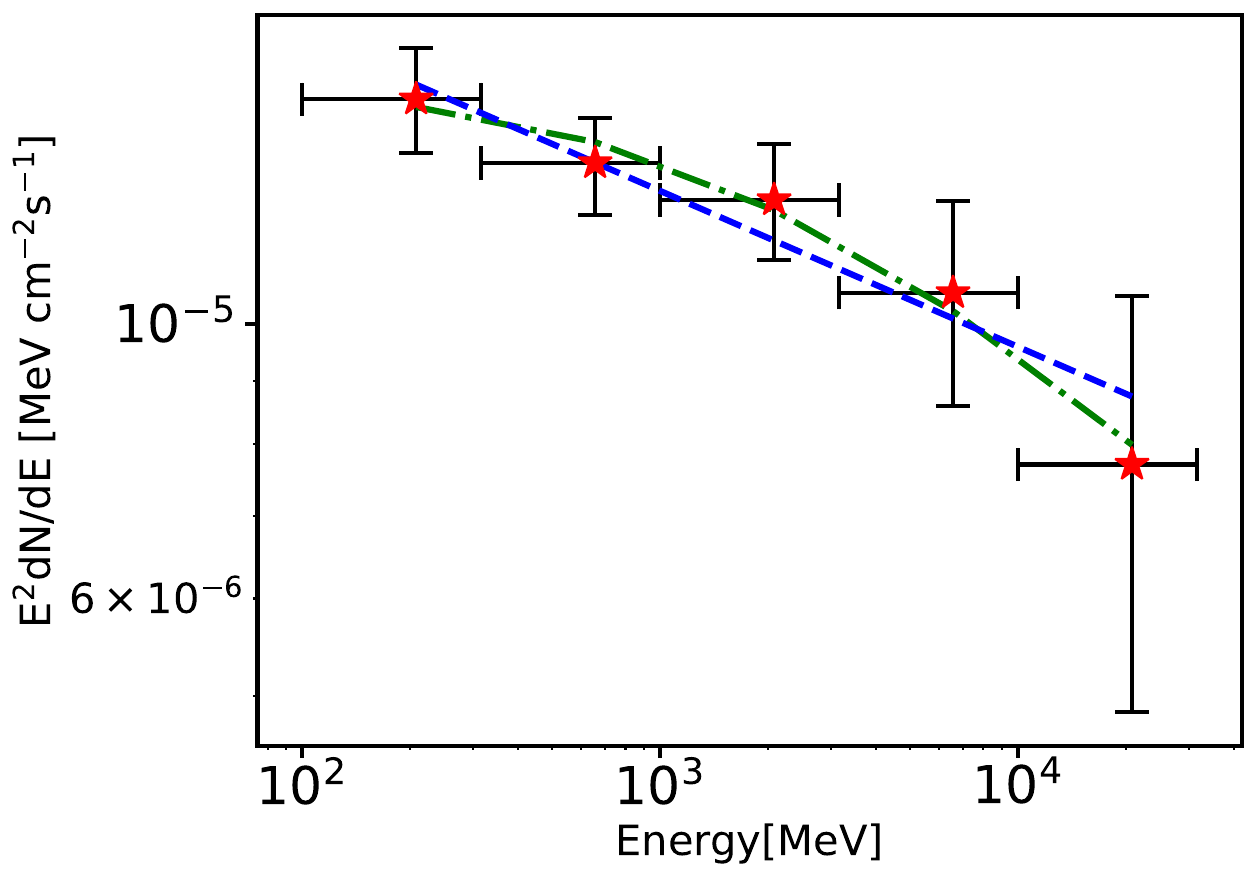}
      }

\hbox{
     \includegraphics[scale=0.27]{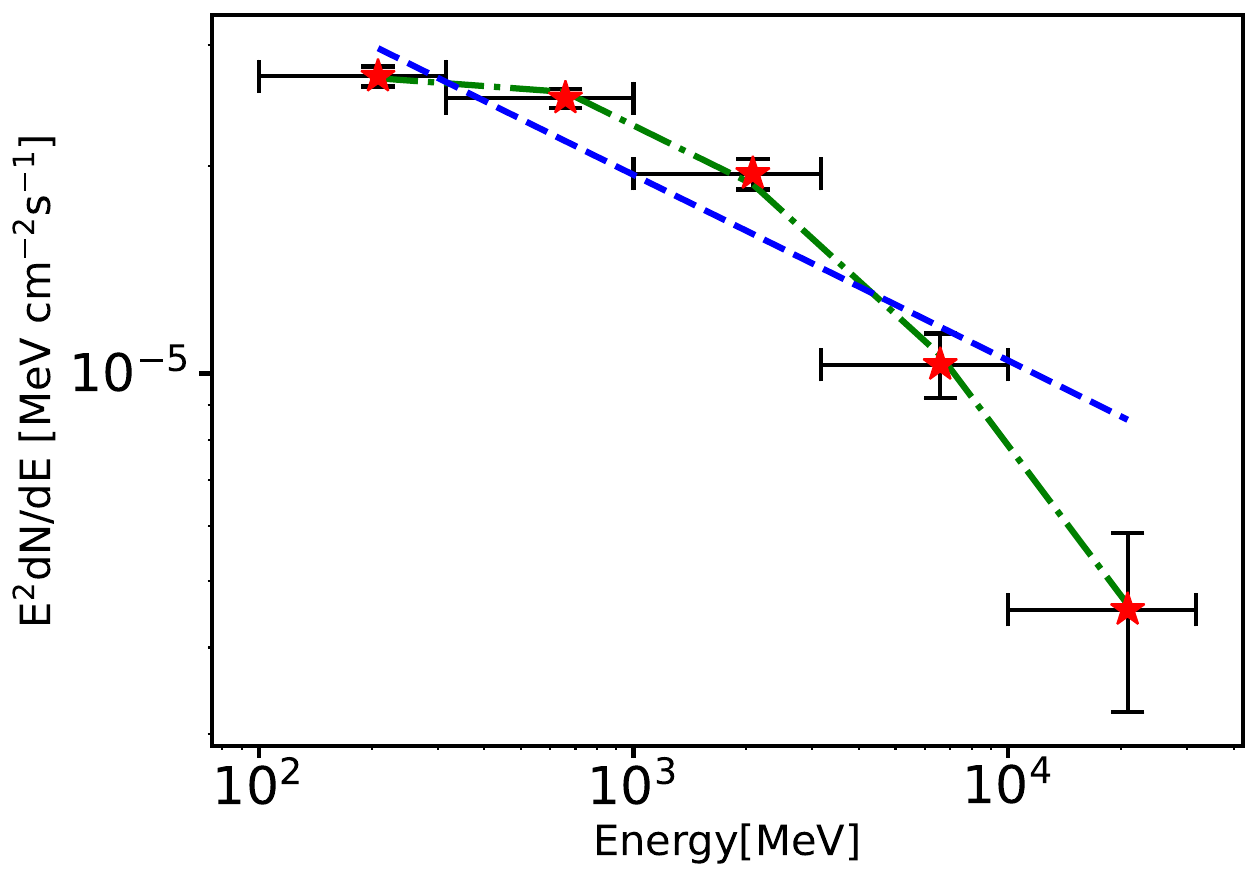}
     \includegraphics[scale=0.27]{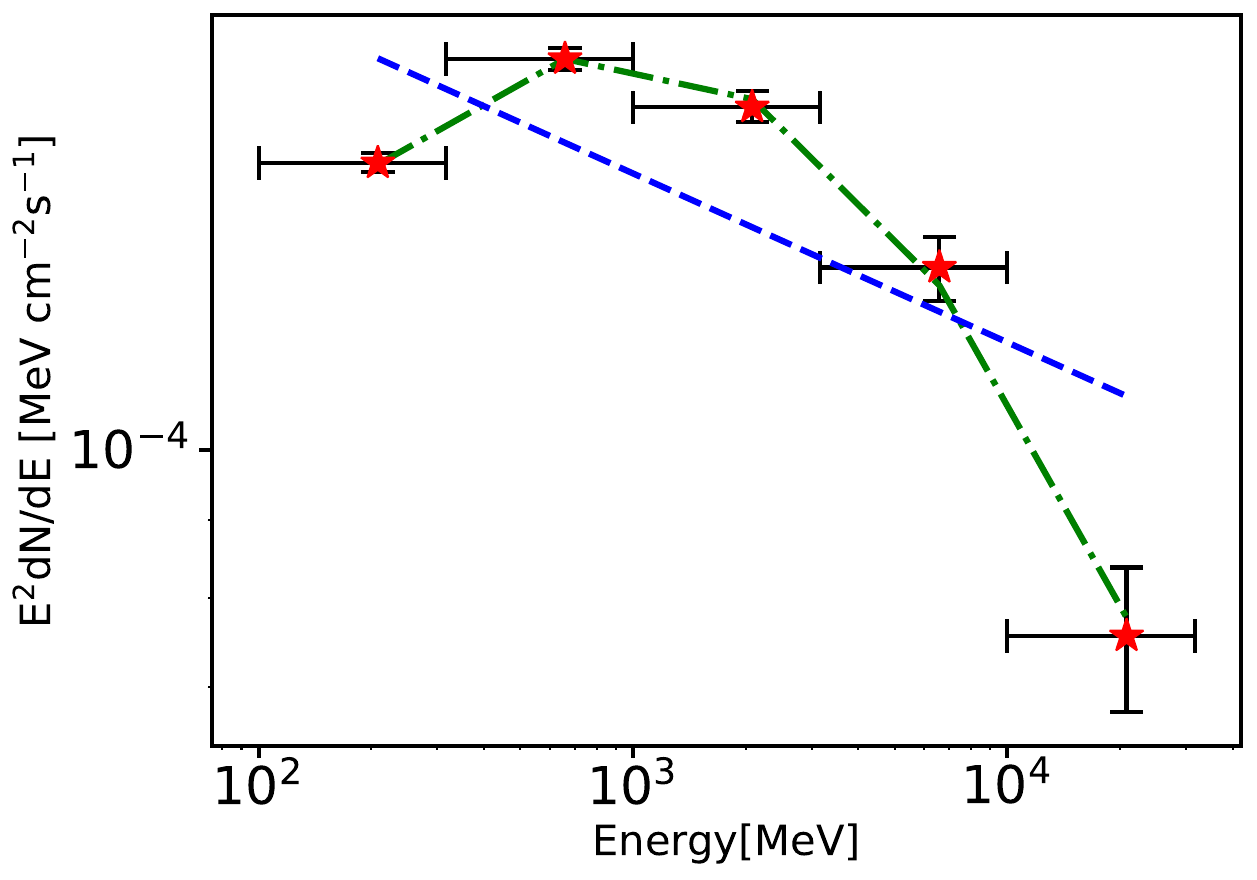}
     \includegraphics[scale=0.27]{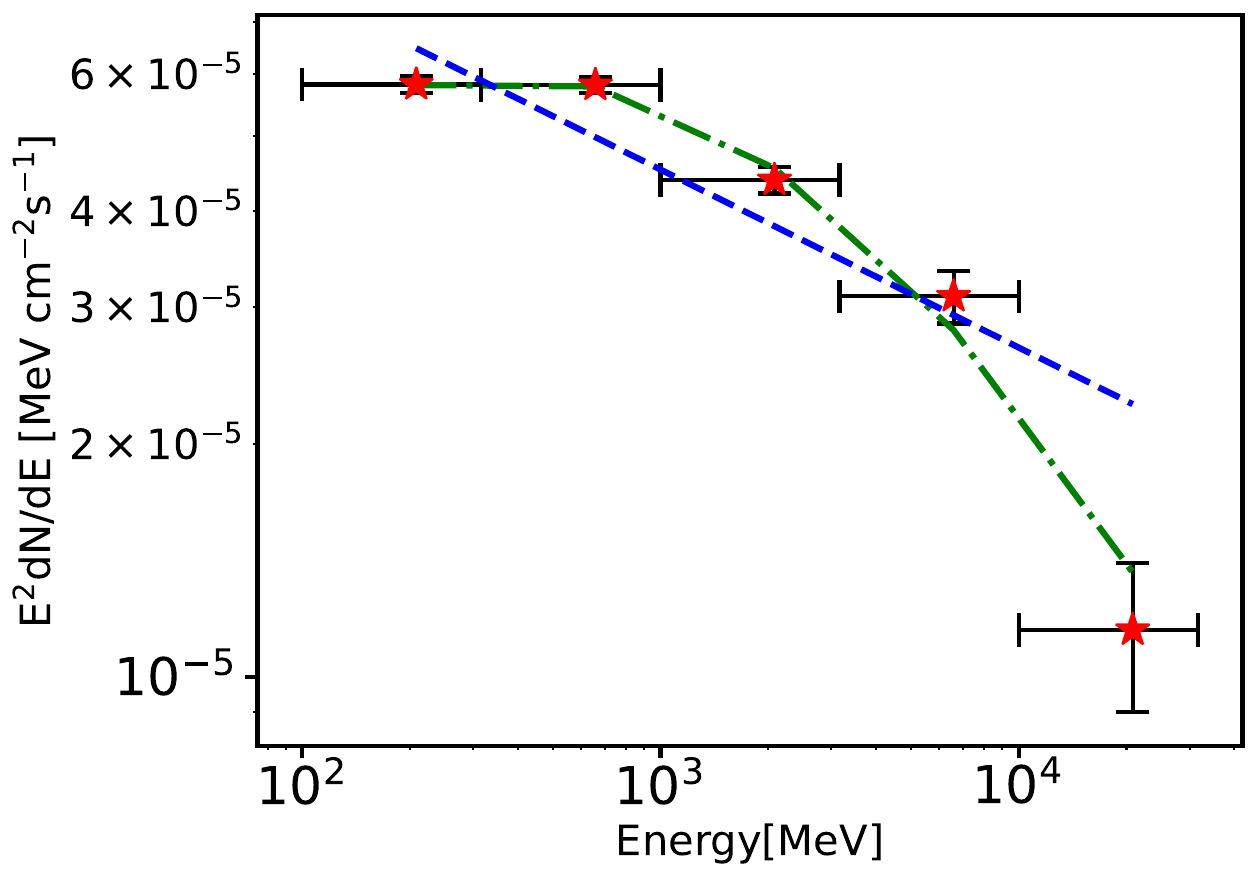}
      }
\hbox{
     \includegraphics[scale=0.27]{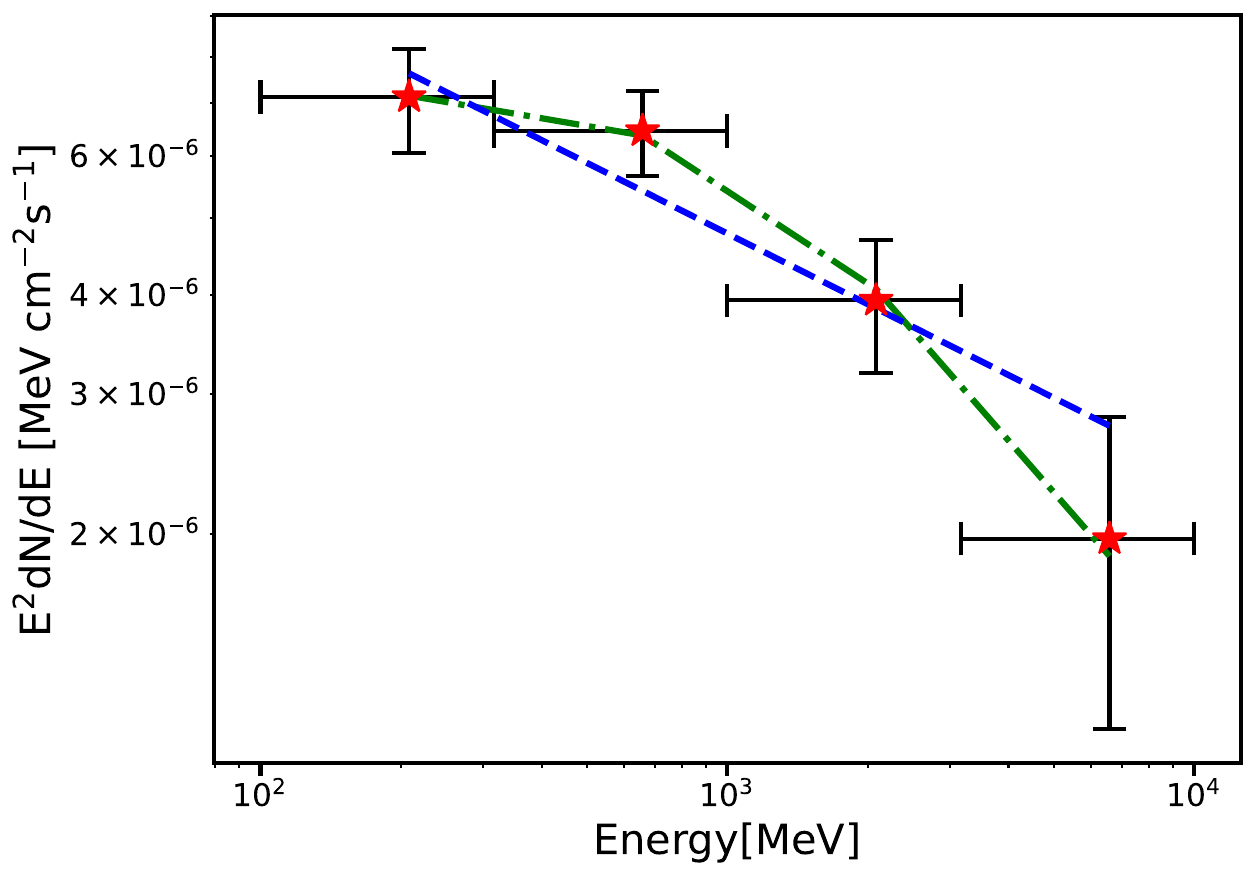}
     \includegraphics[scale=0.27]{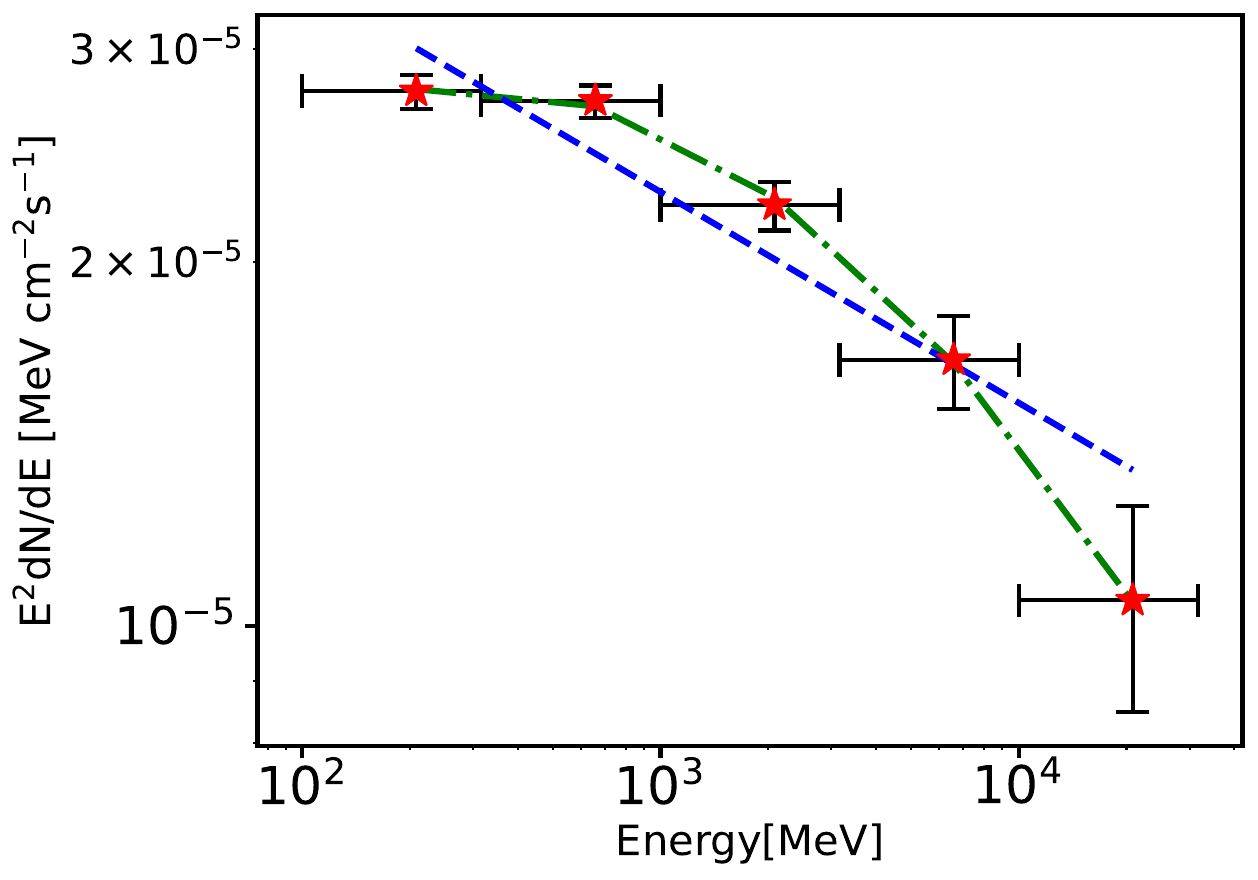}
     \includegraphics[scale=0.27]{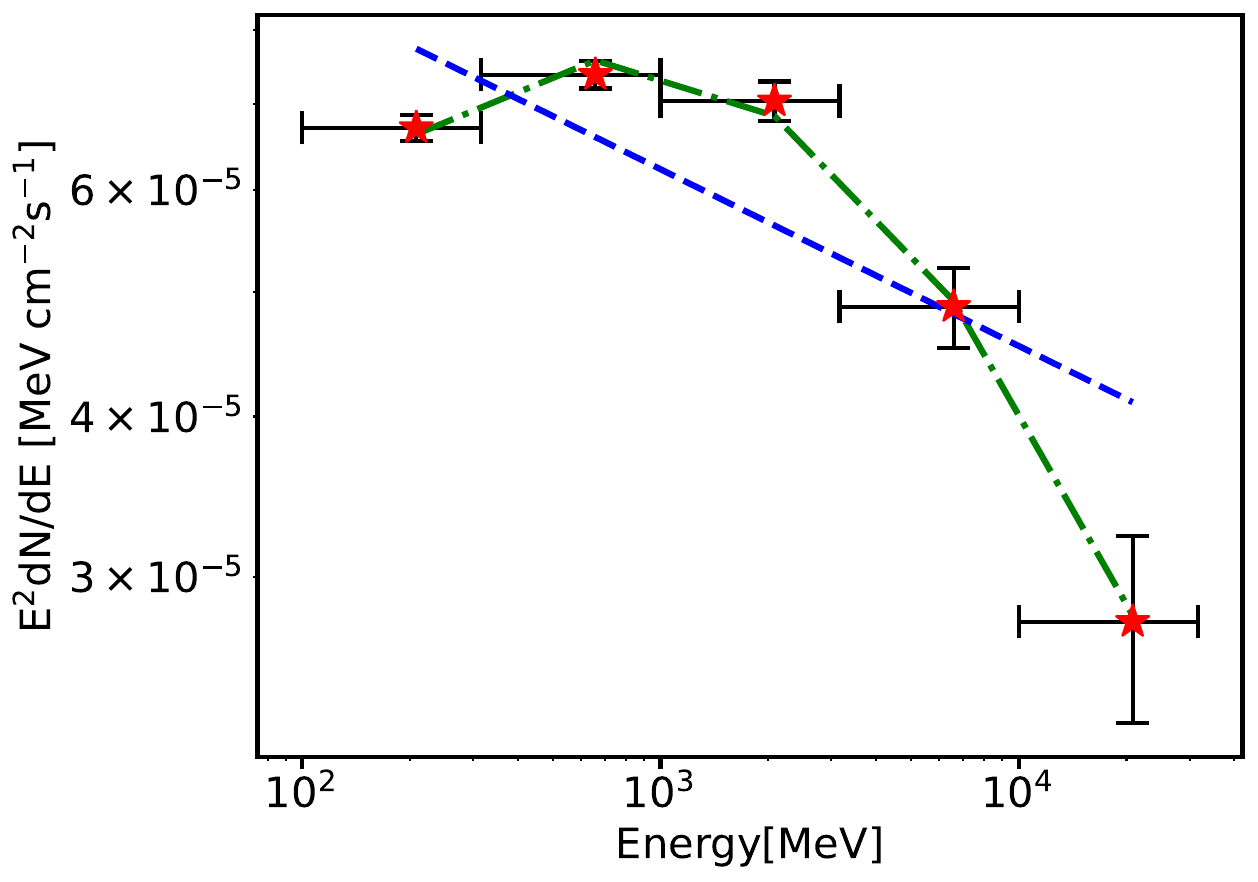}
      }
}

\end{center}
\caption{Simple power law (blue line) and log parabola (green line) fits to the $\gamma$-ray spectra of the sources analyzed in this work during the epochs $E_{1}$ (left), $E_{2}$ (middle) and $E_{3}$ (right). For the top panels to the bottom panels, the sources are PKS 0446+112, TXS 0506+056, PKS 1424$-$418 and PKS 1502+106}
\label{fig-6}
\end{figure*}

We used the maximum likelihood estimator \textit{gtlike} for the spectral analysis likelihood ratio test \citep{1996ApJ...461..396M} to check the PL model (null hypothesis) against the LP model (alternative hypothesis). We calculated the $TS_{curve}$ = 2(log $L_{LP}$$-$log $L_{PL}$) following \cite{2012ApJS..199...31N}. We tested the presence of a significant curvature by setting the condition $TS_{curve}$ $>$ 16.  The $\gamma$-ray spectra along with the model fits for the epochs $E_{1}$, $E_{2}$ and $E_{3}$ for the sources PKS 0446+112, TXS 0506+056, PKS 1424$-$418 and PKS 1502+016 are shown in Fig. \ref{fig-6}. The parameters obtained for both PL and LP model fits are given in Table~\ref{tab:PL_PL}. The $\gamma$-ray spectra for all three epochs of PKS 0446+112 were well fit with a PL model. In the case of TXS 0506+056, epochs $E_{1}$ and $E_{2}$ were fit with a PL model, while epoch $E_{3}$ required the LP model. For PKS 1424$-$418 and PKS 1502+106, all the epochs were well described by a LP model. 

\begin{table*}
\centering
\caption{Results of the PL and LP model fits to the selected epochs of the four sources. Here, $\Gamma$ is the photon index, Flux is the $\gamma$-ray flux value, TS is the test statistics, $\alpha$ is the spectral index, $\beta$ is the measure of curvature in the spectra and $TS_{curve}$ signifies the presence of curvature in the spectra.}
\label{tab:PL_PL}
\resizebox{\textwidth}{!}{%
\begin{tabular}{llccrrrrrrrrr} 
\hline
Source & \multicolumn{1}{c}{Epochs}  & \multicolumn{4}{c}{PL} & \multicolumn{6}{c}{LP} \\ 
     &  &  $\Gamma$  & Flux  & TS  & L$_{PL}$ & $\alpha$  & $\beta$  & Flux  & TS  & L$_{LP}$  & TS$_{curve}$ \\ \hline
  
\multirow{3}{*}{PKS 0446+112} &  $E_{1}$ & $-$2.56  & 2.9$\times$$10^{-11}$ & 102.7 &$-$32727.4 & 2.55 $\pm$0.13 & 0.01 $\pm$ 0.00 & 3.1$\times$$10^{-11}$ & 104.0 & $-$32730.4 &  $-$6 \\
& $E_{2}$ & $-$2.31  & 1.1$\times$$10^{-10}$ & 592.9 & $-$28125.6 & 2.16 $\pm$0.07 & 0.17 $\pm$ 0.04 & 1.2$\times$$10^{-10}$ & 604.6 & $-$28118.8 & 13.6 \\
& $E_{3}$ & $-$2.33 & 8.9$\times$$10^{-11}$ & 721.5 & $-$35483.8 & 2.22 $\pm$0.07 & 0.11 $\pm$ 0.04 & 5.2$\times$$10^{-12}$ & 728.5 & $-$35480.6 & 7.6 \\  \hline

TXS 0506+056 &  $E_{1}$  &  $-$2.16 & 9.9$\times$$10^{-12}$ & 864.9 & $-$36394.3&  2.14$\pm$ 0.05 & 3.7$\times$$10^{-4}$$\pm$ 2.5$\times$$10^{-5}$ & 9.8$\times$$10^{-12}$ & 837.4 & $-$36394.8 &$-$0.8 \\
& $E_{2}$  &  $-$2.12 & 5.6$\times$$10^{-12}$ & 836.5& $-$56832.2&  2.09$\pm$ 0.06 & 0.06 $\pm$ 0.03 & 6.1$\times$$10^{-12}$ & 830.7 & $-$56834.1 & 3.8\\
& $E_{3}$ & $-$2.0 & 2.2$\times$$10^{-11}$ &3635.1 & $-$48360.6&  1.99$\pm$ 0.03 & 0.07 $\pm$ 0.02 & 2.4$\times$$10^{-11}$ & 3629.2 & $-$48348.7 & 23.8 \\\hline

\multirow{4}{*}{PKS 1424$-$418} &  $E_{1}$ &  $-$2.25 & 4.6$\times$$10^{-11}$ & 4513.2 & $-$81341.7 &  2.21 $\pm$ 0.02 & 0.11 $\pm$ 0.02 & 5.5$\times$$10^{-11}$ & 4704.5& $-$81298.6 & 86.2\\
& $E_{2}$ &  $-$1.86 & 1.0$\times$$10^{-10}$ & 23884.6 & $-$54766.7 &  1.99 $\pm$ 0.01 & 0.09 $\pm$ 0.01 & 3.9$\times$$10^{-10}$ & 31919.6 & $-$53817.0 & 1899.4\\
& $E_{3}$ &  $-$2.21 & 9.9$\times$$10^{-11}$ & 10111.1 & $-$67345.2 &  2.16 $\pm$0.02 & 0.09 $\pm$ 0.01 & 1.2$\times$$10^{-10}$ &10314.7 & $-$67299.3 & 91.8\\ \hline

\multirow{3}{*}{PKS 1502+106} &  $E_{1}$ &  $-$2.06 & 1.0$\times$$10^{-11}$ & 201.5 & $-$28964.8 & 2.38 $\pm$0.09 & 0.02 $\pm$ 0.01 & 2.4$\times$$10^{-11}$ & 698.6 & $-$29004.9 & 238.8 \\
& $E_{2}$ &  $-$1.37 & 1.0$\times$$10^{-11}$ & 3550.5 & $-$56113.1 & 2.07 $\pm$0.03 & 0.08 $\pm$ 0.01 & 1.1$\times$$10^{-10}$ & 6629.5 & $-$55081.3 & 2063.6 \\ 
& $E_{3}$ & $-$1.09 & 1.0$\times$$10^{-11}$ & 5274.9 & $-$42376.6  & 1.96 $\pm$0.02 & 0.09 $\pm$ 0.02 & 3.1$\times$$10^{-10}$ & 13108.4 & $-$39469.9 &  5813.4  \\ \hline
\end{tabular}%
}
\end{table*}

\begin{figure*}
\begin{center}
\vbox{
      \hbox{
\includegraphics[scale=0.42,angle=270]{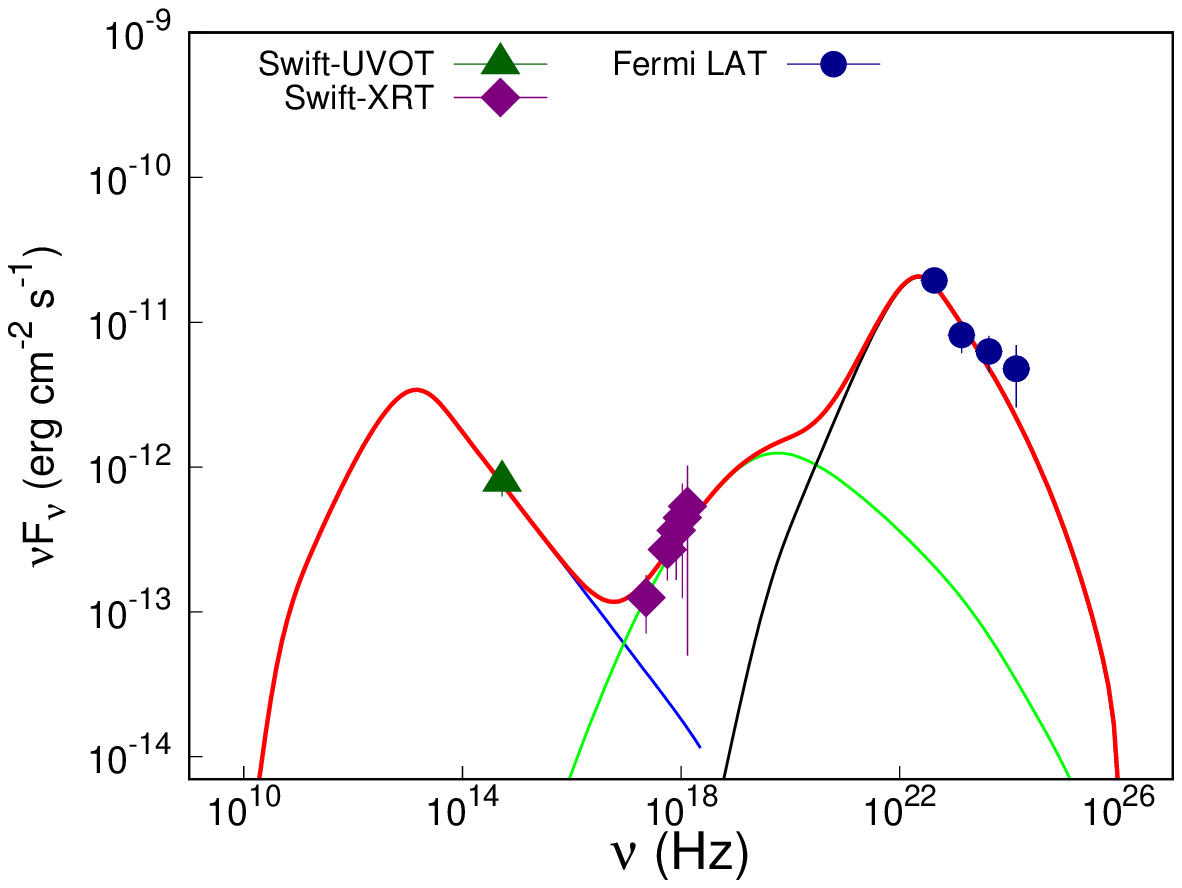}
\includegraphics[scale=0.42,angle=270]{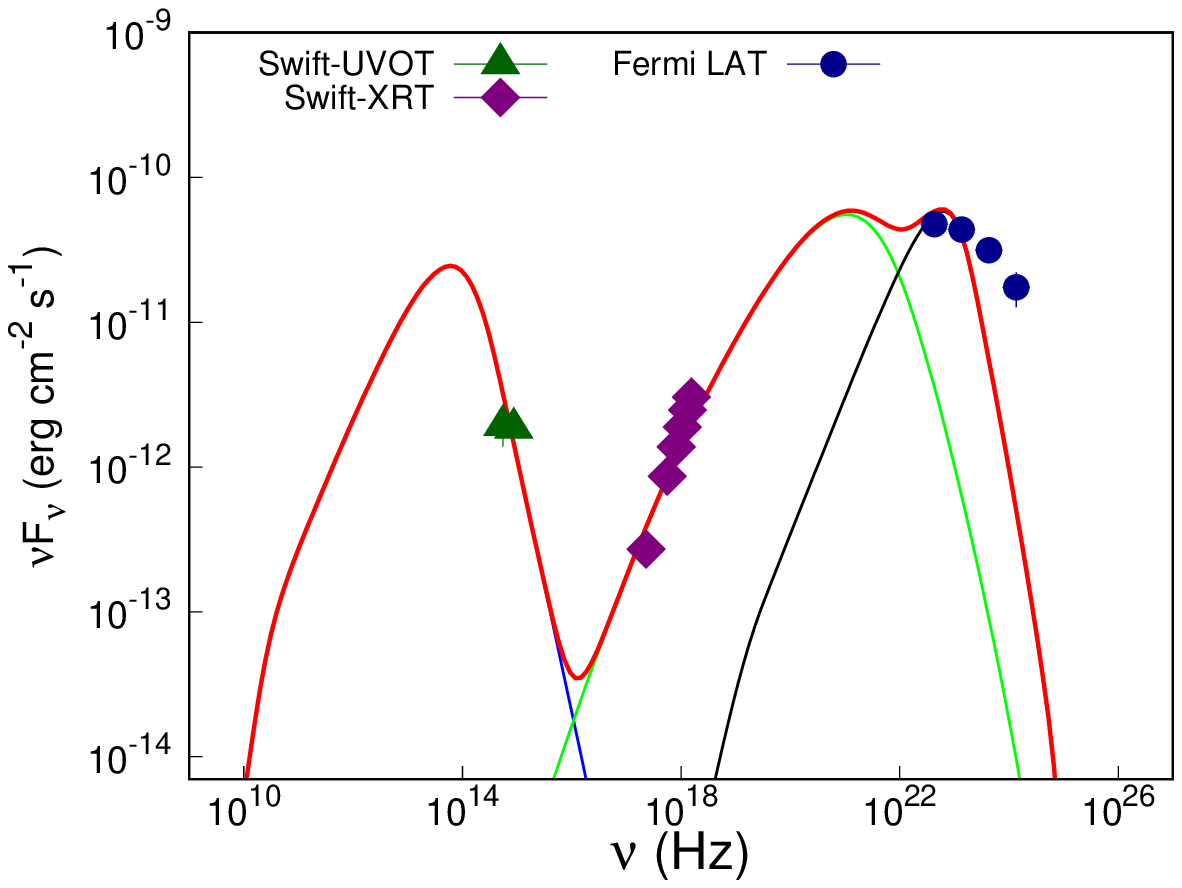}
           }
     \hbox{
\includegraphics[scale=0.42,angle=270]{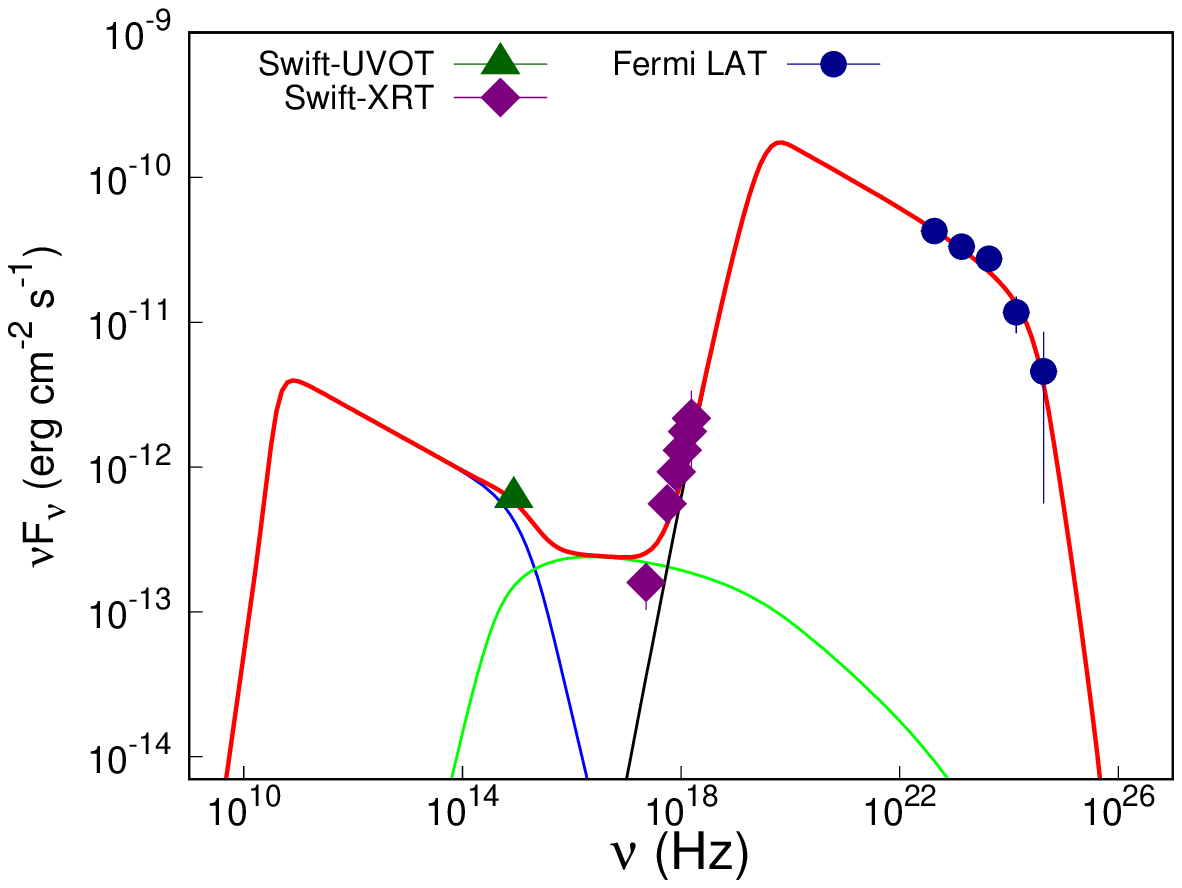}
\includegraphics[scale=0.35, angle=270]{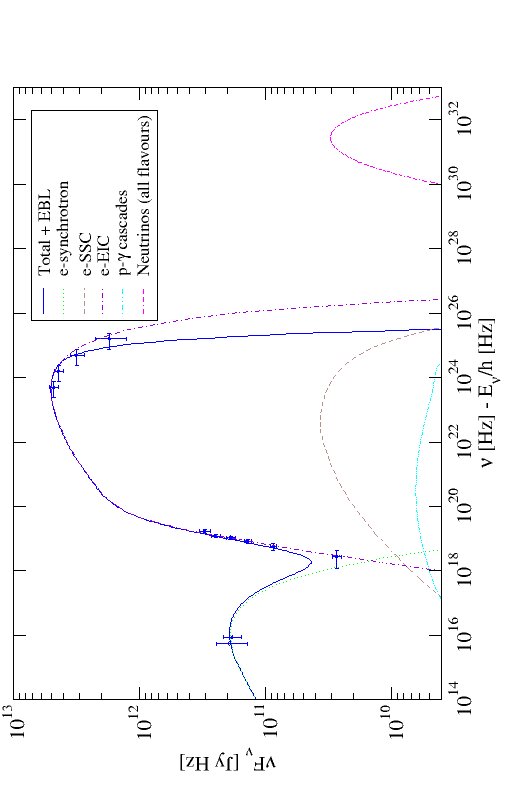}
           }
     }
\end{center}
\caption{Broadband SEDs along with the one zone leptonic emission model fits for the epochs $E_{1}$ (top left panel), $E_{2}$ (top right panel) and $E_{3}$ (bottom left panel) of  PKS 0446+112. In all the fits, blue line refers to the synchrotron model, the green line refers to the SSC process and the black line refers to the EC process. The red line is the sum of all the three components. The bottom-right panel shows the hadronic fit to the SED of the epoch $E_{2}$. }
\label{fig-7}
\end{figure*}

\begin{figure}
\begin{center}
\vbox{
\includegraphics[scale=0.42,angle=270]{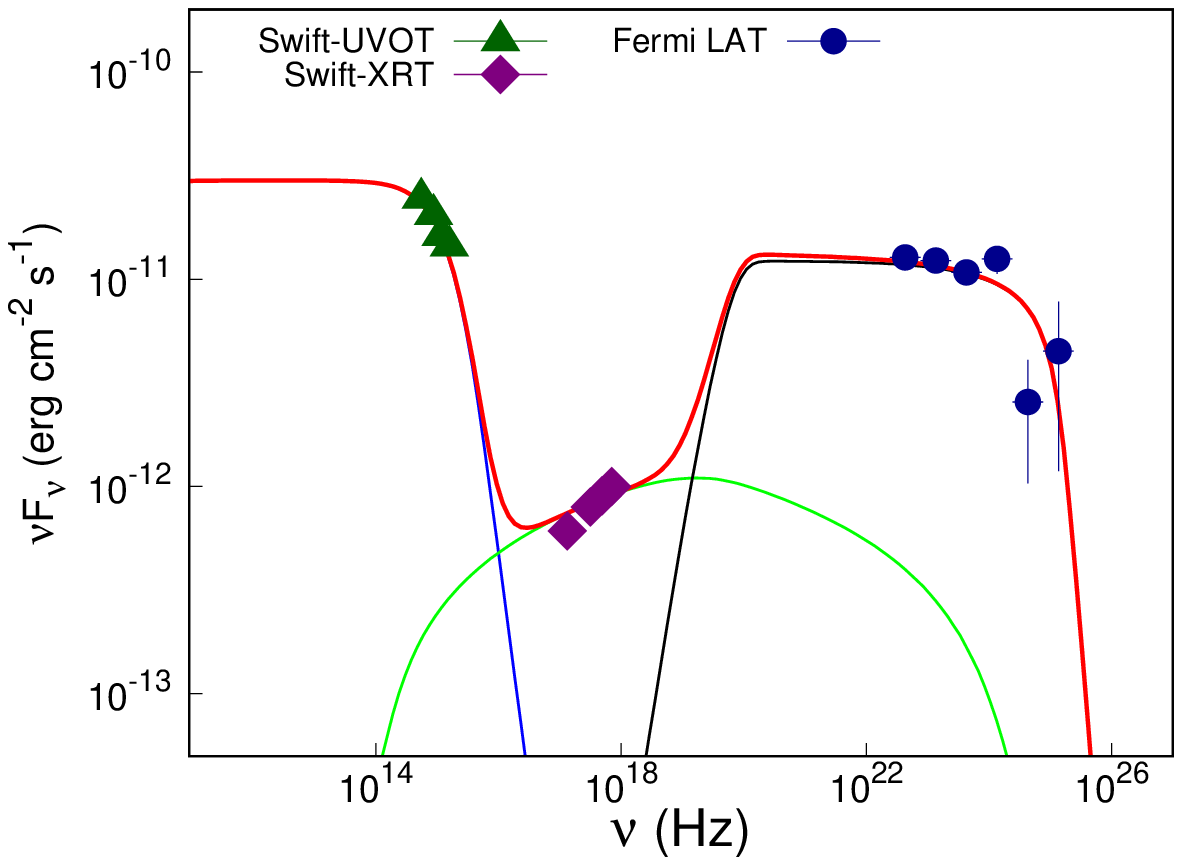}
\includegraphics[scale=0.42,angle=270]{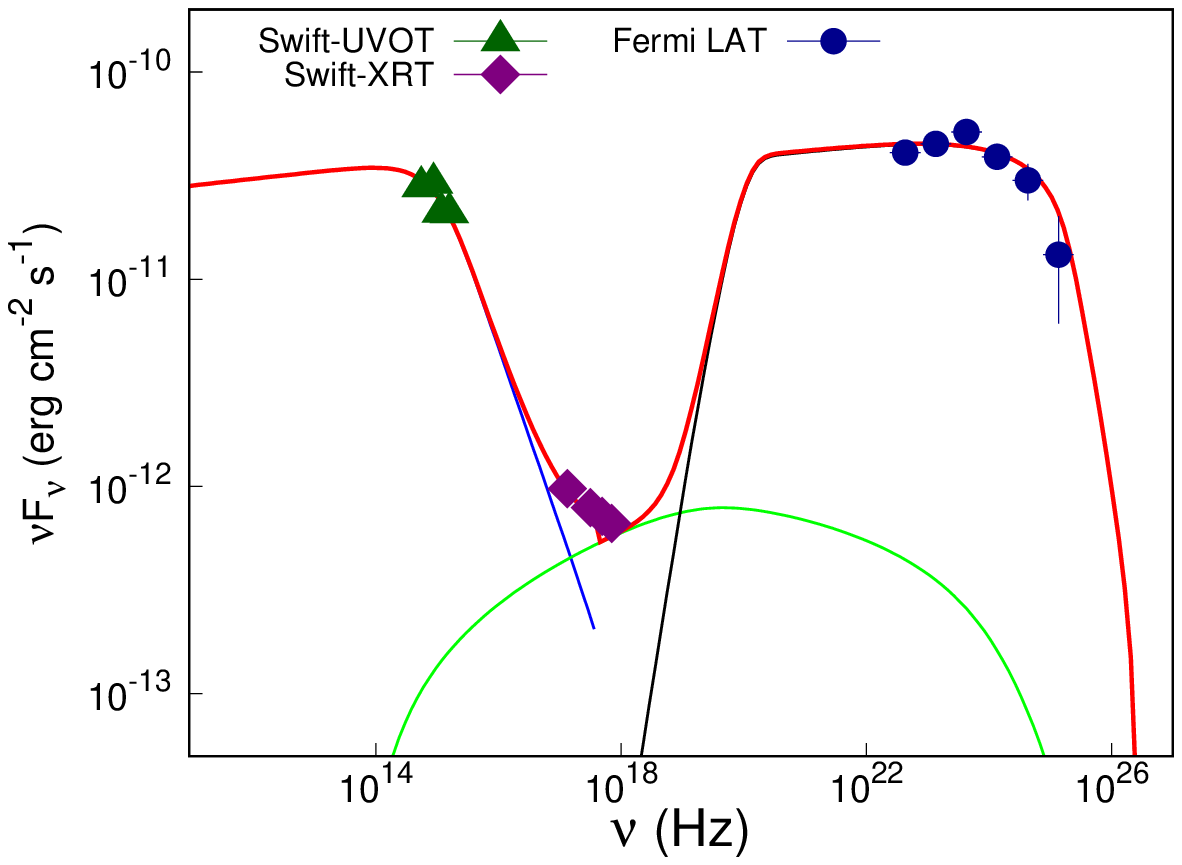}
\includegraphics[scale=0.42,angle=270]{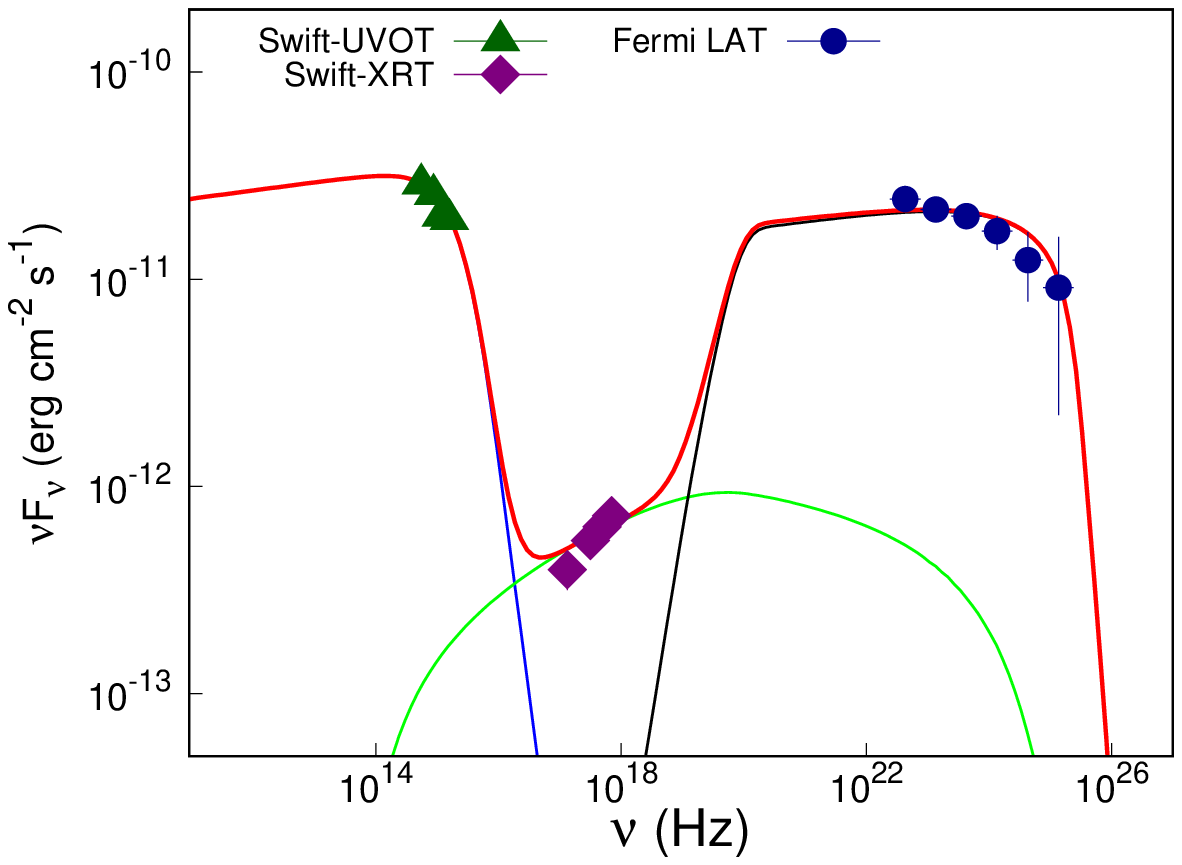}
}
\end{center}

\caption{Broadband SEDs along with the one zone leptonic emission model fits for the epochs $E_{1}$ (top panel), $E_{2}$ (middle panel) and $E_{3}$ (bottom panel) of TXS 056+056. In all the fits, blue line refers to the synchrotron model, the green line refers to the SSC process and the black line refers to the EC process. The red line is the sum of all the three components.}
\label{fig-8}
\end{figure}

\begin{figure}
\begin{center}
\vbox{
\includegraphics[scale=0.42,angle=270]{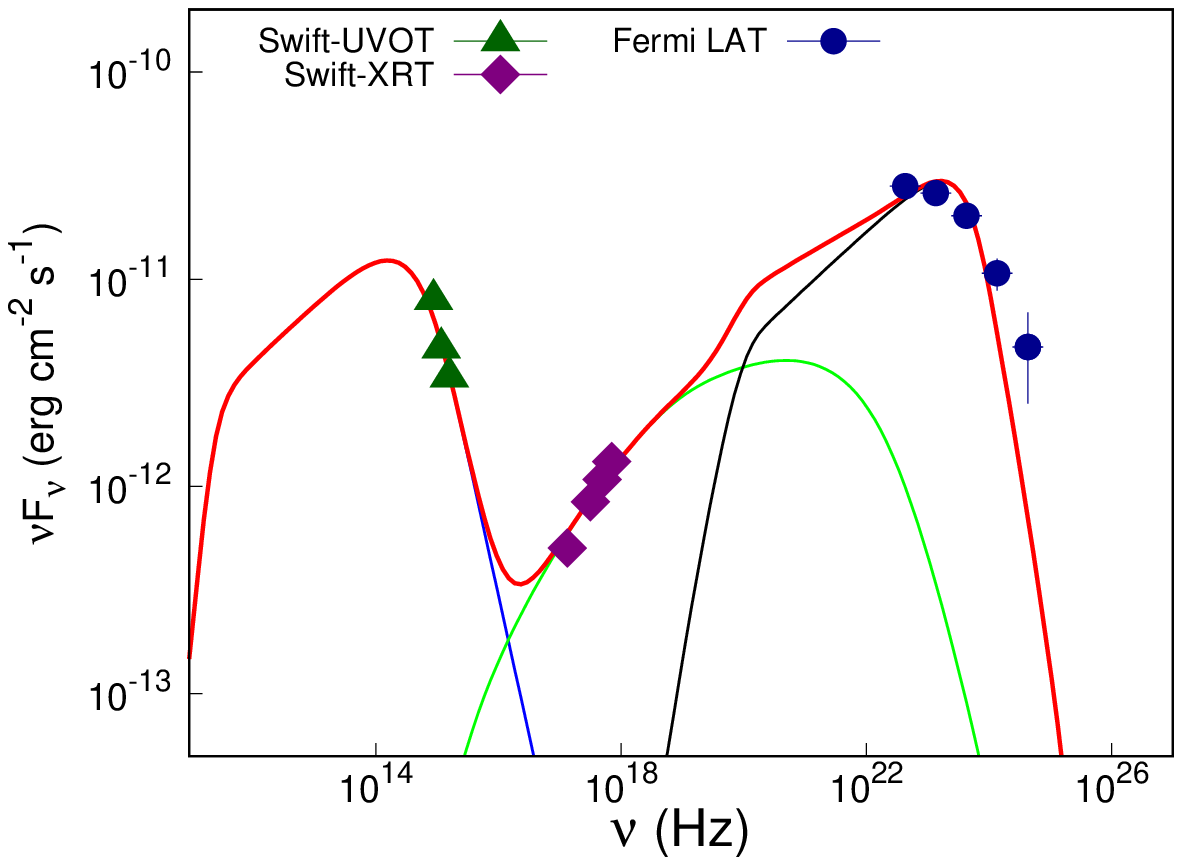}
\includegraphics[scale=0.42,angle=270]{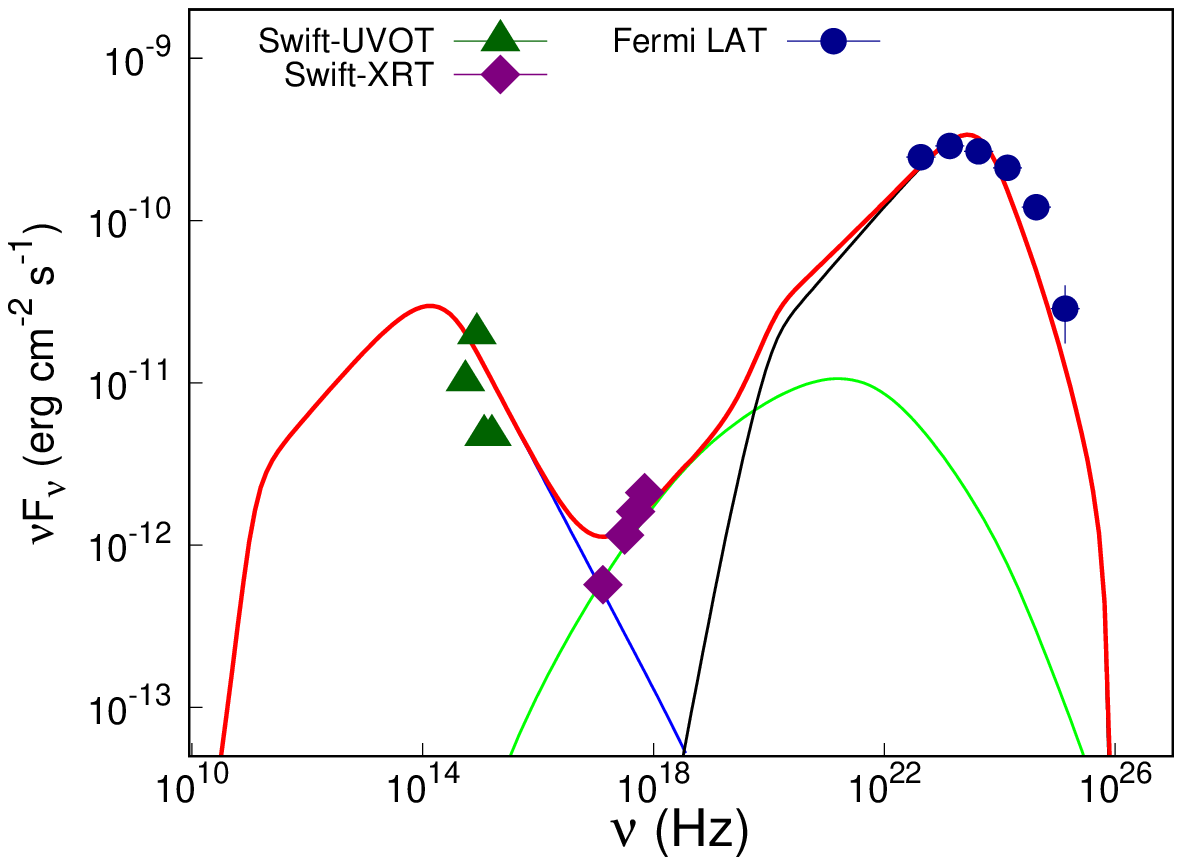}
\includegraphics[scale=0.42,angle=270]{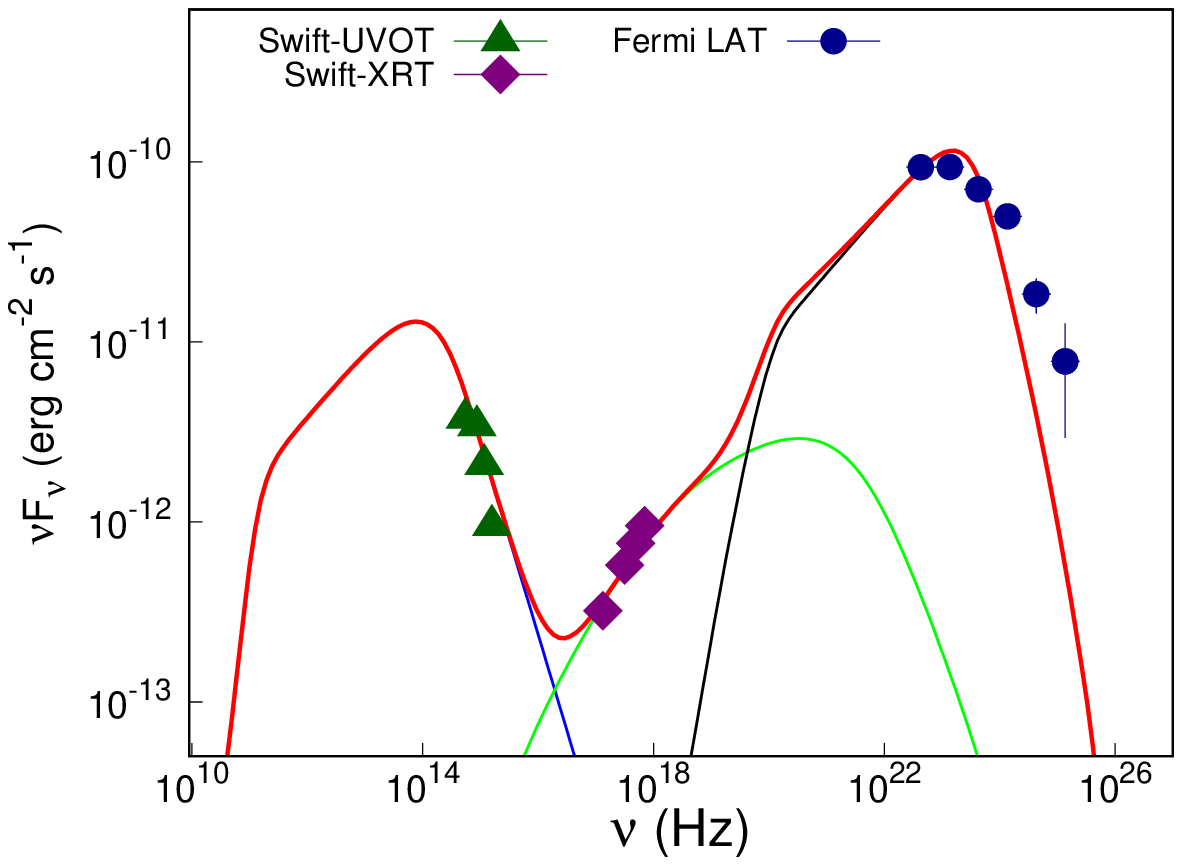}
}
\end{center}
\caption{Broadband SEDs along with the one zone leptonic emission model fits for the epochs $E_{1}$ (top panel), $E_{2}$ (middle panel), and $E_{3}$ (bottom panel) of PKS 1424$-$418. In all the fits, the blue line refers to the synchrotron model, the green line refers to the SSC process, and the black line refers to the EC process. The red line is the sum of all the three components.}
\label{fig-9}
\end{figure}

\begin{figure*}
\begin{center}
\vbox{
     \hbox{
\includegraphics[scale=0.42,angle=270]{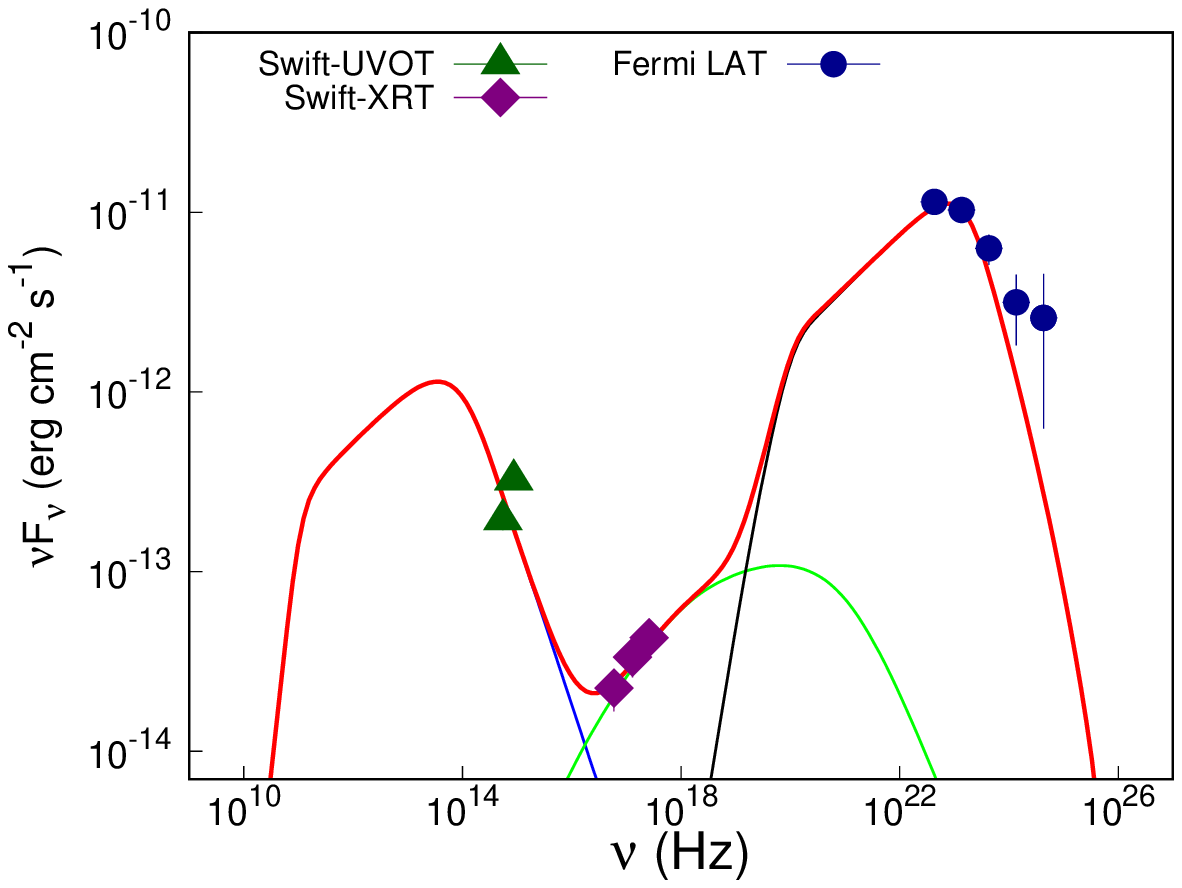}
\includegraphics[scale=0.42,angle=270]{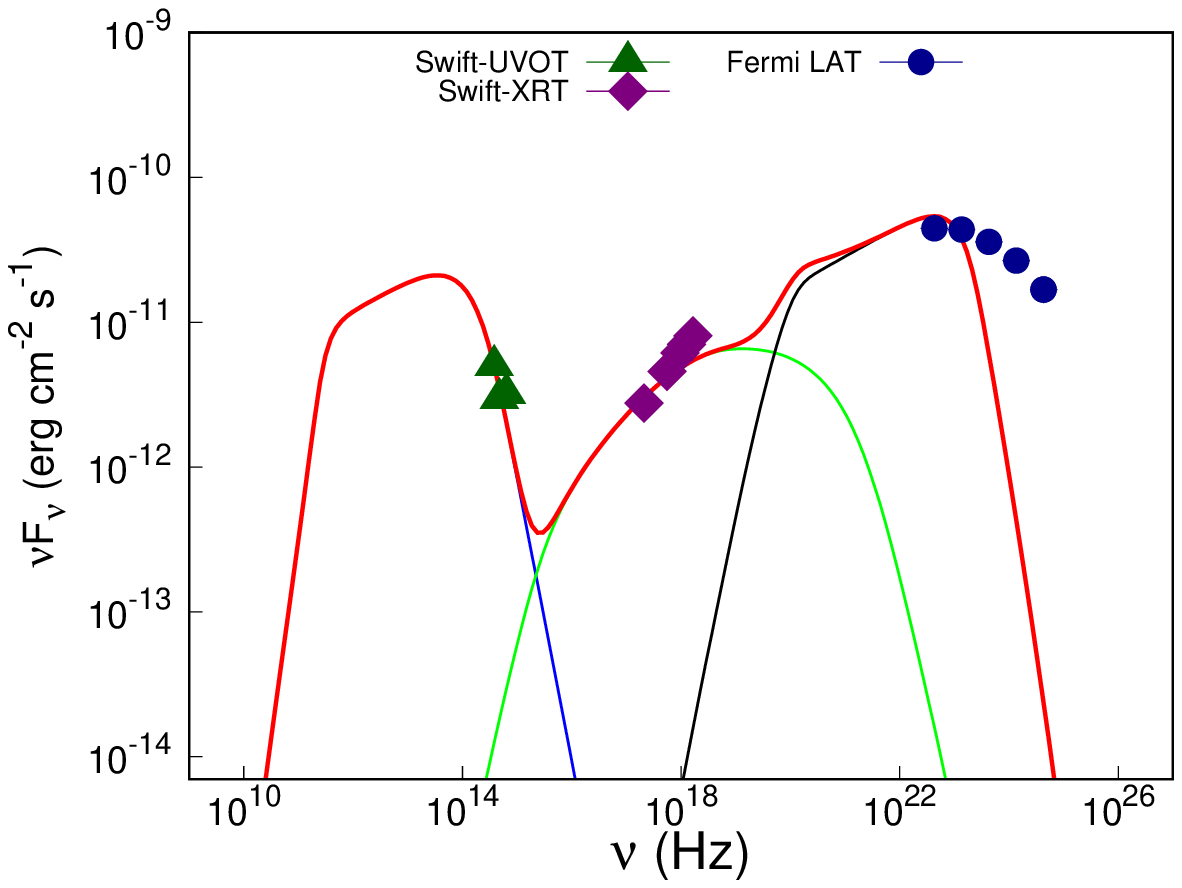}
          }
     \hbox{
\includegraphics[scale=0.42,angle=270]{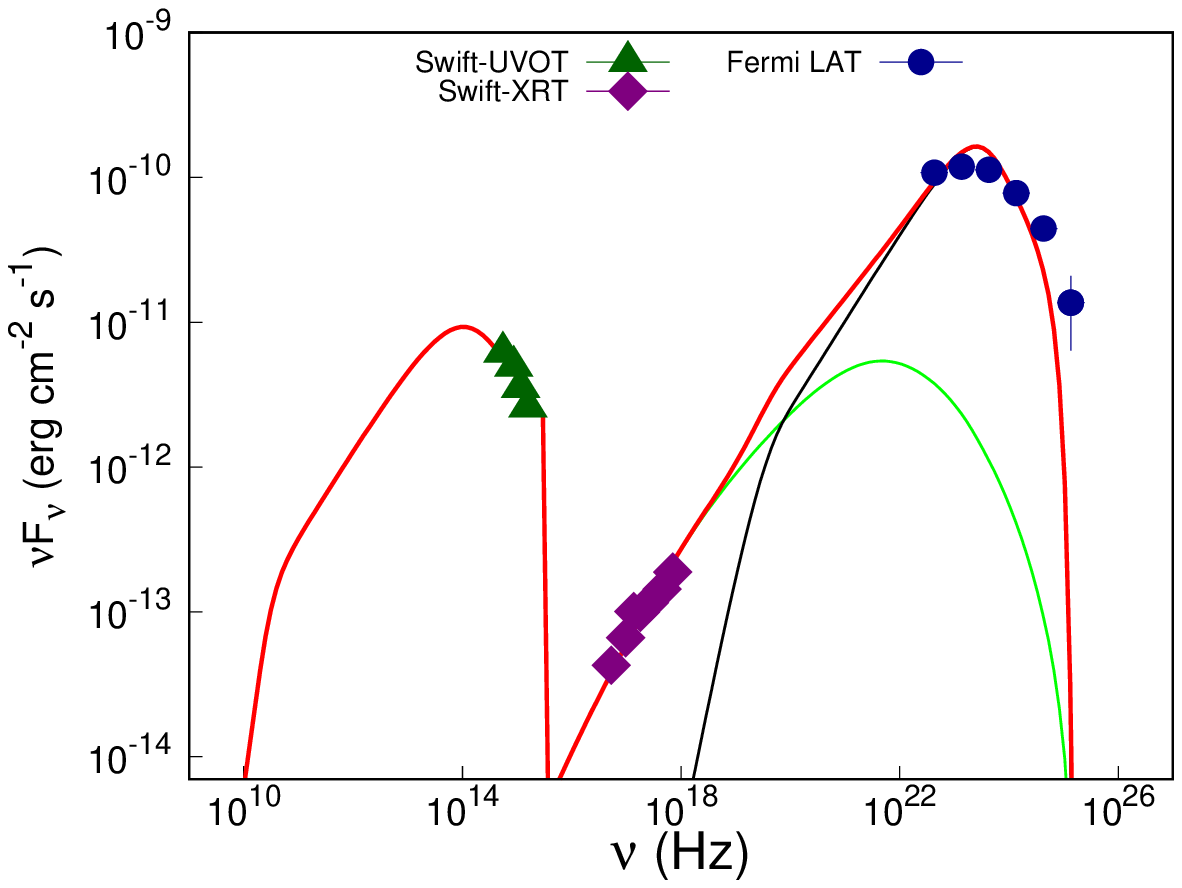}
\includegraphics[scale=0.35, angle=270]{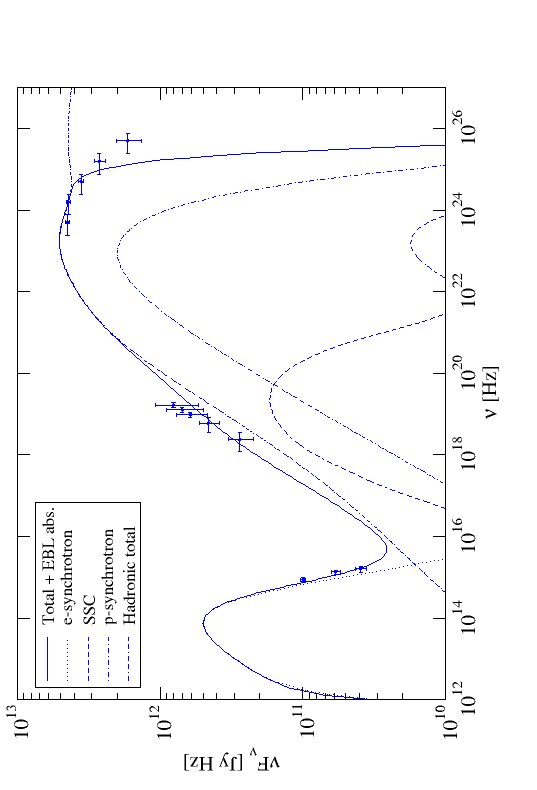}
         }       
}
\end{center}
\caption{Broadband SEDs along with the one zone leptonic emission model fits for the epochs $E_{1}$ (top left panel), $E_{2}$ (top right panel) and $E_{3}$ (bottom left panel) of  PKS 1502+106. In all the fits, blue line refers to the synchrotron model, the green line refers to the SSC process and the black line refers to the EC process. The red line is the sum of all the three components. The bottom-right panel shows the hadronic fit to the SED of the epoch $E_{2}$.  }
\label{fig-10}
\end{figure*}

\subsection{Broadband SED analysis}
\subsubsection{One-zone Leptonic modeling}
We also carried out broadband SED analysis of all the three epochs for each of the sources. For this we used the one-zone leptonic framework described in \cite{2012MNRAS.419.1660S}. This model assumes the emission to originate from a spherical region of radius $R$, moving down the jet relativistically with Lorentz factor ($\Gamma$) and filled with a tangled magnetic field. The jet is inclined at an angle $\theta$ with the line of sight and hence the observed emission will be boosted by the Doppler factor $\delta$. The emission region is filled with a non-thermal electron distribution following a broken power-law with indices $p$ and $q$, described as,
\begin{align} \label{eq:broken}
	N(\gamma)\,d\gamma = \left\{
\begin{array}{ll}
	K\,\gamma^{-p}\,d\gamma&\textrm{for}\quad \mbox {~$\gamma_{\rm min}<\gamma<\gamma_b$~} \\
	K\,\gamma_b^{q-p}\gamma^{-q}\,d\gamma&\textrm{for}\quad \mbox {~$\gamma_b<\gamma<\gamma_{\rm max}$~}
\end{array}
\right.
\end{align} 
where, $\gamma$ is the electron Lorentz factor and $\gamma_b$ corresponds to the break Lorentz factor. The electrons radiate via synchrotron emission and inverse Compton scattering through either SSC or EC processes. We assumed the emission region to be beyond the broad line region (BLR) but within the IR torus. Hence, due to relativistic boosting the dominant photon fields will be the IR dust emission which is chosen as a blackbody at 1000 K. The emissivity due to these emission processes was computed numerically under steady-state, and the flux received on Earth is used to reproduce the observed spectrum. The best fit SEDs using this model for PKS 0446+112, TXS 0506+056,  PKS 1424$-$418 and  PKS 1502+106 are shown  in Figs. \ref{fig-7}, \ref{fig-8}, \ref{fig-9}, and \ref{fig-10}, respectively. The results of the SED analysis are given in Table~\ref{table-7}.

\begin{table*}
	\centering
	\caption{Results of the broadband SED analysis carried out for the selected epochs of PKS 1424$-$418, PKS 1502+106, PKS 0446+112 and TXS 0506+056. The parameters $p$ and $q$ are the low and high energy power law indices of the electron distribution, $\gamma_b$ is the the break energy, R is the size of the emission blob in cm, $\Gamma$ is the bulk Lorentz factor and $B$ is the magnetic field in Gauss. In our modeling, we considered a minimum electron Lorentz factor of $\gamma_{\rm min,e} = 50$ and a maximum of $\gamma_{\rm max,e} = 9.5 \times 10^{5}$.}
\label{table-7}
\begin{tabular}{llccrrrr} 
	\hline
Source & Epoch & $p$ & $q$ & $\gamma_b$ &  log (R) (cm) & $\Gamma$  & $B$ (G)\\
		\hline
PKS 0446+112    & $E_{1}$ &  1.16 & 3.99 & 1156.3 & 16.99 & 12.54 & 0.34 \\
                & $E_{2}$ & 1.1   & 6.49 & 3890.6 & 17.42 & 50    & 0.34 \\ 
		          & $E_{3}$ & 3.42  & 6.18 & 18331  & 18.09 & 11.97 & 0.12\\ \hline

TXS 0506+056    & $E_{1}$ &  3.0  & 7.5 & 25867   & 18.49 & 9.02  & 0.03\\ 
		          & $E_{2}$ & 2.93  & 4.7 & 26153   & 18.6  & 11.91 & 0.02\\
		          & $E_{3}$ &  2.92 & 7.5 & 33971.5 & 18.54 & 10.05 & 0.02\\ \hline
        
PKS 1424$-$418  & $E_{1}$  & 2.29 & 6.84 & 3874.6  & 16.68 & 50    & 0.68  \\
		& $E_{2}$  & 2.17 & 4.36 & 3681.2  & 17.07 & 28.61 & 0.32  \\
		& $E_{3}$  & 2.25 & 5.13 & 2933.6  & 16.92 & 28.63 & 0.37  \\ \hline
        
PKS 1502+106    & $E_{1}$ & 2.38 & 4.84 & 2215.2 & 16.72 & 28.9 & 0.36\\
		& $E_{2}$ &  2.57 & 6.9 & 2000 & 16.81 & 29.6 & 0.70\\ 
		& $E_{3}$ &  1.77 & 7.00 & 4702.78 & 16.88 & 29.63 & 0.35\\ \hline
	\end{tabular}
\end{table*}

\subsubsection{Hadronic SED modeling}

For the epochs in which the simple one-zone leptonic model failed to reproduce the observed high-energy emission (epoch $E_{2}$ of PKS~0446+112 and epoch $E_{2}$ of PKS~1502+106), we employed the stationary hadronic model described in \cite{2013ApJ...768...54B}. The model assumes a spherical emission region of radius $R$, moving relativistically with bulk Lorentz factor $\Gamma$ and Doppler factor $\delta$, and permeated by a tangled magnetic field $B$, similar to the leptonic framework.

In addition to a non-thermal electron population, the emission region contains relativistic protons injected with a power-law energy distribution between $\gamma_{p,\min}$ and $\gamma_{p,\max}$. Particle escape is parameterized by an escape timescale $t_{\rm esc} = \eta_{\rm esc} R/c$, while the particle distributions are evolved under steady-state conditions accounting for radiative particle cooling and interactions.

The relativistic electrons radiate through SSC processes, while the high-energy component is dominated by proton synchrotron radiation and photo-hadronic ($p\gamma$) interactions. The proton synchrotron emission is calculated using the standard synchrotron formalism,

\begin{equation}
P_{\rm syn}(\nu) \propto 
\int N_p(\gamma_p)\,
F\!\left(\frac{\nu}{\nu_c(\gamma_p)}\right)
\, d\gamma_p ,
\end{equation}
where $N_p(\gamma_p)$ is the proton energy distribution and $\nu_c(\gamma_p) = \frac{3 e B}{4 \pi m_p c}\gamma_p^2$ is the critical proton synchrotron frequency. The $p\gamma$ interactions are treated using the semi-analytical approach of \cite{2013ApJ...768...54B}, based on the templates of \cite{2008PhRvD..78c4013K}, which provide the spectra of secondary photons, neutrinos, and electron–positron pairs. The model self-consistently accounts for secondary emission processes, including photons from $\pi^0$ decay, synchrotron radiation from secondary $e^\pm$ produced in charged pion and muon decays, and synchrotron-supported pair cascades initiated by internal $\gamma\gamma$ absorption. This framework was applied to epoch $E_{2}$ of PKS~0446+112 and PKS~1502+106, where the purely leptonic model could not adequately reproduce the observed $\gamma$-ray flux. The resulting best-fit SEDs are shown in Figs.~\ref{fig-7} and \ref{fig-10}, and the corresponding parameters are listed in 
Table~ \ref{tab:model_parameters}.

For PKS~0446+112, the hadronic modeling yields results that are qualitatively consistent with the findings of \cite{2025arXiv251107643K}. The very hard X-ray spectrum strongly constrains the possible contribution from hadronic processes, effectively ruling out a dominant proton-synchrotron component in the observed SED. Instead, the broadband emission is primarily electron-dominated, with only a subdominant hadronic contribution at the highest energies. The derived model parameters exhibit near-equipartition between particle and magnetic energy densities, with $L_{B}/L_{e} \sim 1$, indicating a physically well-balanced jet configuration. 

The predicted neutrino flux from photo-hadronic interactions is found to be very low, consistent with the absence of a significant excess reported by the IceCube Neutrino Observatory apart from the single neutrino alert. Overall, while a lepto-hadronic scenario can reproduce the observed SED, the hard X-ray spectrum implies that the emission is predominantly leptonic in nature during this epoch.

For PKS~1502+106, a satisfactory reproduction of the high-energy component required a hybrid scenario in which the low- and intermediate-energy emission is dominated by synchrotron and SSC processes, while the $\gamma$-ray emission is primarily contributed by hadronic interactions. Although this configuration provides a reasonable fit to the observed SED, the derived physical parameters are rather extreme. In particular, the required proton kinetic power is highly super-Eddington, and the electron energy distribution is characterized by a very hard spectral index ($\sim 5$), which is uncommon in standard shock-acceleration scenarios. These factors suggest that, while the lepto-hadronic model can formally reproduce the observed spectrum, the required physical conditions may be challenging from an energetic and theoretical standpoint.

\begin{table*}
\centering
\caption{Results of hadronic model fits to epoch $E_{2}$ for PKS~0446+11 and PKS~1502+106.}
\begin{tabular}{lcc}
\hline
Parameter & PKS~0446+11 & PKS~1502+106 \\
\hline
Minimum Lorentz factor ($\gamma_{\rm min}$) & $50$ & $7\times10^{2}$ \\
Maximum Lorentz factor ($\gamma_{\rm max}$) & $5\times10^{4}$ & $1\times10^{4}$ \\
Electron spectral index ($\alpha_{e}$) & $1.7$ & $5.0$ \\
Magnetic field ($B$ [G]) & $1.1$ & $3$ \\
Blob radius ($R$ [cm]) & $3\times10^{16}$ & $3.5\times10^{15}$ \\
Bulk Lorentz factor ($\Gamma$) & $20$ & $35$ \\
Observing angle ($\theta_{\rm obs}$ [$^\circ$]) & $1$ & $1$ \\
Accretion disk luminosity ($L$) & $1\times10^{46}$ & $1\times10^{46}$ \\
Black hole mass [$M_\odot$] & $5\times10^{8}$ & $5\times10^{8}$ \\
$E_{p,\min}$ [GeV] & $1$ & $1$ \\
$E_{p,\max}$ [GeV] & $2.5\times10^{9}$ & $1\times10^{9}$ \\
Proton spectral index ($\alpha_p$) & $1.9$ & $2.0$ \\
IR temperature$T_{\rm ext}$ [K] & $1\times10^{3}$ & -- \\
IR photon field$u_{\rm ext}$ [erg cm$^{-3}$] & $1.5\times10^{-3}$ & -- \\
$L_e$ [erg s$^{-1}$] & $1.6\times10^{45}$ & $2.4\times10^{44}$ \\
$L_B$ [erg s$^{-1}$] & $1.6\times10^{45}$ & $5.1\times10^{44}$ \\
$L_p$ [erg s$^{-1}$] & $1.0\times10^{45}$ & $9.0\times10^{49}$ \\
$L_B/L_e$ & $1.02$ & $2.11$ \\
$L_B/L_p$ & $1.63$ & $5.63\times10^{-6}$ \\
$L_e/L_p$ & $1.60$ & $2.67\times10^{-6}$ \\

\hline

\end{tabular}
\label{tab:model_parameters}
\end{table*}

\section{Discussion and Summary}
Blazars are now increasingly found to be associated with neutrinos detected by IceCube. These observations are therefore beginning to put constraints on the physical processes that happen in the relativistic jets of these sources. The neutrino detection along with simultaneous/near simultaneous followup observations over a range of wavelengths, have now enabled better understanding of the broadband SEDs of blazars which also involves development of theories or models involving leptons, hadrons and photons \citep{2022MNRAS.509.2102G,2024AJ....168..289S,2024A&A...683A.225S}. These multi-messenger observations will help in a better understanding of their relativistic jets. In this work, we carried out such a study on four neutrino blazars namely PKS 0446+112, TXS 0506+056, PKS 1424$-$418 and PKS 1502+106. The results are summarized below

\begin{enumerate}
\item Of the  four sources studied in this work, we found a temporal correlation between neutrino detection and flaring $\gamma$-ray activity in three sources, namely, PKS 0446+112, TXS 0506+056 and PKS 1424$-$418. However, we found no temporal correlation between neutrino detection and $\gamma$-ray flaring activity in the source PKS 1502+106. Similarly, during the neutrino flare detected in the year 2014/15,   in TXS 0506+056, no flaring activity  in $\gamma$-rays was noticed \citep{2020ApJ...893..162F}. Therefore, neutrino detection does not necessarily correspond to flaring $\gamma$-ray activity. These findings consistent with earlier results \citep{2021JCAP...10..082O, 2023ApJ...954...75A} underscore the complexity of blazar emission processes and indicate that neutrino emission in blazars may not necessarily coincide with their flaring $\gamma$-ray activity states.

\item All the four sources are found to be variable in $\gamma$-rays. Concentrating on the brightest $\gamma$-ray  activity of the sources (which also corresponds to the epoch of neutrino detection in PKS 0446+112, TXS 0506+056 and PKS 1424$-$418) we found flux doubling/halving time scale of 4.70 hrs, 9.24 hrs, 30.76 hrs and 15.42 hrs in the sources PKS 0446+112, TXS 0506+056, PKS 1424$-$418 and PKS 1502+106, respectively. From these observed flux doubling/halving timescale, we found the size of the emission region as 3.70 $\times$ 10$^{15}$ cm for PKS 0446+112, 7.47 $\times$ 10$^{16}$ cm for TXS 0506+056, 2.63 $\times$ 10$^{16}$ cm for PKS 1424$-$418 and 7.04 $\times$ 10$^{15}$ cm for PKS 1502+106. These values are found to lie in the range mentioned by \cite{2023ApJS..268...23F} through an analysis of thousands of {\it Fermi} blazars.  This points to a compact emission region within the relativistic jets of these sources. The difference in emission region sizes from rapid variability and SED modeling is expected. Variability traces compact sub-structures responsible for fast flares, while SED modeling reflects a larger, averaged region producing the total flux. Thus, the two estimates probe different physical scales within the jet.

\item The location of the $\gamma$-ray emitting region (relative to the central black hole) was estimated using the observed flux doubling/halving timescales together with the bulk Lorentz factors derived from the SED modeling \citep{2010MNRAS.405L..94T}.  
\begin{equation}
r_{\rm em} \lesssim \frac{2 c \Gamma^{2} \tau}{1+z}
\end{equation}
We obtained distances of $r_{\rm em} \approx 8.1 \times 10^{17}\,\mathrm{cm}$ for PKS 0446+112, $2.1 \times 10^{17}\,\mathrm{cm}$ for TXS 0506+056, $2.2 \times 10^{18}\,\mathrm{cm}$ for PKS 1424$-$418, and $1.0 \times 10^{18}\,\mathrm{cm}$ for PKS 1502+106. These values suggest that the location of the emission region is beyond the typical BLR radius ($10^{16}$$-$$10^{17}$ cm; \citealt{1996MNRAS.280...67G}), but within the dusty torus ($10^{18}$$-$$10^{19}$ cm). Hence, the dominant external Compton emission mechanism will be the EC scattering of IR photons.

\item We also investigated the spectral variations of the sources in all epochs where both soft and hard $\gamma$-ray light curves could be generated. For PKS 1424$-$418 and PKS 1502+106, the diagrams suggest a tendency for spectral hardening with increasing total brightness, although the scatter in the data points prevents firm conclusions. For TXS 0506+056, the data indicate a possible softer-when-brighter behaviour.

\item For all the sources, we investigated the nature of the $\gamma$-ray spectra in three epochs, one corresponding to the epoch of neutrino detection, the second one corresponding to a quiescent activity of the sources and the third one corresponding to a flaring state of the sources. The spectra were well fit with a power law for the source PKS 0446+112. For the source TXS 0506+056, power-law model fits the spectra well during epochs $E_{1}$ and $E_{2}$, while during epoch $E_{3}$, log parabola model fits the spectrum well compared to the power-law model. For the sources PKS 1424$-$418 and PKS 1502+016 a log parabola model was found to well explain the spectra.

\item For the source PKS 1424$-$418, all three epochs analyzed in this study are well modeled in the leptonic scenario. However, \cite{2017ApJ...843..109G}, who investigated the source during different activity phases concluded that a lepto-hadronic model with a sub dominant hadronic component could explain its SED . A more recent investigation by \cite{2024A&A...681A.119R} also found the SED of PKS 1424$-$418 to be compatible with a leptonic model in agreement with the results presented in this work.

\item For PKS 1502+106, the broadband SEDs during both the quiescent state (Epoch $E_{1}$) and the flaring state (Epoch $E_{3}$) were successfully reproduced with a leptonic emission model incorporating synchrotron, SSC, and EC processes. However, during epoch $E_{2}$, which corresponds to the time of the neutrino detection and a faint $\gamma$-ray flux state, an additional hadronic component was required to explain the broadband emission (see bottom right panel of Fig.~\ref{fig-10}). Within the lepto-hadronic framework, the high-energy emission during this epoch is primarily attributed to proton synchrotron radiation and $p\gamma$ interactions. Although this scenario provides a satisfactory fit to the observed SED, it requires extreme physical conditions, including a highly super-Eddington proton kinetic power and an unusually hard electron energy distribution. These energetic requirements suggest that, while the hadronic contribution can account for the observed spectrum, the implied jet parameters are challenging from a physical standpoint.

The short variability timescale and the probable detection of a neutrino during this epoch further support the presence of both leptonic and hadronic emission mechanisms. Modeling of the broadband SED of PKS 1502+106 across three activity states, including the neutrino detection epoch, was previously carried out by \cite{2021ApJ...912...54R}. We note here that our identification of flaring and quiescent epochs was based on visual examination and the requirement of multi-band data. In contrast, \cite{2021ApJ...912...54R} identified flaring periods using Bayesian block analysis and defined the quiescent state based on flux measurements during a low-activity period. They found that hadronic models can account for the multi-wavelength emission during all three states as well as the observed neutrino signal during its quiescent state.A similar scenario is observed in another neutrino source, PKS~0735+178, for which SED modeling indicates that the quiescent state required a photo-meson process to explain the emission, whereas the flaring state, which coincided with the neutrino detection epoch, could be adequately explained by a one-zone leptonic model \citep{2023MNRAS.519.1396S, 2023ApJ...954...70A, 2024MNRAS.527.8746P, 
2024MNRAS.529.3503B, 2025A&A...695A.266O}.

\item For the source PKS~0446+112, while the SEDs of the quiescent and flaring episodes are well reproduced by purely leptonic processes, the epoch of neutrino detection (Epoch $E_{2}$) required an additional hadronic component to account for the observed high-energy emission. The very hard X-ray spectrum during this epoch strongly constrains the contribution from proton synchrotron radiation, implying that the emission remains largely electron-dominated with only a subdominant hadronic contribution. The derived model parameters are close to equipartition ($L_{B}/L_{e} \sim 1$), indicating a physically plausible jet configuration. The predicted neutrino flux in this scenario is very low, consistent with the absence of a significant excess reported by IceCube apart from the single neutrino alert.

\item For TXS 0506+056, all the three epochs analyzed in this work including the one coinciding with the 2017 IceCube-170922A are well explained by a purely leptonic emission scenario. Previously, \cite{2020ApJ...891..115P} modeled the broadband SED of the source using contemporaneous data in optical, X-ray and $\gamma$-ray energies across four epochs, including the period 2014-2015, during which a neutrino flare with a significance of ~3.5$\sigma$ was reported from the direction of TXS 0506+056 \citep{2018Sci...361..147I}. They found the SEDs are consistent with a hybrid scenario, where the $\gamma$-rays are produced by EC processes and the observed neutrinos arise from photo-meson processes of co-accelerated protons. The difference in the preferred emission scenarios across epochs, purely leptonic in some cases (this work), and hybrid leptonic process in other epochs \citep{2020ApJ...891..115P} may be due to (a) the varying dominance of leptonic and hadronic components at different epochs, (b) the emission originating from different regions during different epochs and/or (c)
complex physical processes not fully captured by current models.
\end{enumerate}

This work underscores the importance of coordinated multi-wavelength and multi-messenger observations to unravel the complexities of blazar jets. It is anticipated that more neutrino blazars will be detected in the future with the current as well as future neutrino experiments. As of now, we are aware of  two sources, namely, TXS 0506+056 and PKS 1502+106 which were in a low $\gamma$-ray flux state at the time of IceCube neutrino detections. With an increasing number of detected neutrino blazars, it is likely that more sources exhibiting similar behaviour will be identified. Timing, spectral and SED model fits to a large sample of neutrino blazars, will further help to refine our understanding of the physical mechanisms of these enigmatic sources.

\section*{Acknowledgements}
We thank the referee for constructive comments and suggestions, which helped to improve the quality of this manuscript. Athira M Bharathan acknowledges the Department of Science and Technology (DST) for the INSPIRE Fellowship (IF200255). AMB also thanks the Center for Research, CHRIST (Deemed to be University) and Centre for Space research (CSR), North-west University for their support during the course of this work. This work makes use of archival $\gamma$-ray data from the Fermi Science Support Center (FSSC) and \textit{Swift}-XRT/UVOT data from the High Energy Astrophysics Science Archive Research Center (HEASARC).

\section*{DATA AVAILABILITY}
All the data used here for analysis is publicly available and the results are incorporated in the paper.


\bibliographystyle{mnras}





\newpage
\bsp	
\label{lastpage}

\bibliography{main}

\end{document}